\documentclass[11pt]{article}
\usepackage{amsmath,amsthm}

\newcommand{\R}{{\mathbf R}} \newcommand{\N}{{\mathbf N}}
\newcommand{\K}{{\mathbf K}} \newcommand{\Z}{{\mathbf Z}}
  \def\C{{\mathbf C}}
\newcommand{\Prm}{{\mathbf P}}
\newcommand{\wt}{\widetilde }
\newcommand{\e}{\varepsilon }
\newcommand{\E}{{\mathbf E}}
\renewcommand{\epsilon}{\varepsilon } 
 
\renewcommand{\rho}{\varrho } 
 
\renewcommand{\phi}{\varphi }

\def\rs{\right>}
\def\lg{\left|}

\newtheorem{theorem}{Theorem}
\newtheorem{lemma}{Lemma}
\newtheorem{corollary}{Corollary}

\begin {document}
 \title{Quantum Summation with an Application to Integration}

\author {S.\ Heinrich\\
Fachbereich Informatik\\
Universit\"at Kaiserslautern\\
D-67653 Kaiserslautern, Germany\\
e-mail: heinrich@informatik.uni-kl.de}
   
\date{}
\maketitle

\date{}
\maketitle
\begin{abstract}
We study summation of sequences and integration in the quantum model of 
computation. We develop quantum algorithms for computing the mean of sequences 
which satisfy a $p$-summability condition and for integration of functions from Lebesgue spaces
$L_p\big([0,1]^d\big)$ and  analyze their convergence rates. We also prove lower bounds which show
that the proposed algorithms are, in many cases, optimal within the setting of quantum computing.
This extends recent results of Brassard, H{\o}yer, Mosca, and Tapp (2000) on computing the mean for bounded sequences
and complements results of Novak (2001) on integration of functions from H\"older classes.\\
 
\end{abstract}
\section{Introduction}
Quantum algorithms and complexity are by now well studied for various discrete problems.
This includes such milestones as Shor's (1994) factorization and Gro\-ver's (1996) search algorithm. Much less
is understood about numerical problems, computational problems of analysis. These problems are typically
defined on a continuum and/or take values in a continuum, such as the field of real or complex 
numbers, domains in finite dimensional vector spaces or even infinite dimensional 
normed spaces  like function spaces.

First results related to this direction concern the counting problem \linebreak (Boyer, 
Brassard, H{\o}yer, and Tapp, 1998) and the computation of the mean (Grover, 1998,
Brassard, H{\o}yer, Mosca, and Tapp, 2000) of finite sequences which satisfy a uniform 
bound (e.g.\ whose elements belong to the interval
$[0,1]$). Matching 
lower bounds were obtained by Nayak and Wu (1999) using the polynomial method of 
Beals, Buhrman, Cleve, and Mosca (1998).
Abrams and Williams (1999) proposed 
certain quantum algorithms for integration. Novak (2001) was the first to provide quantum 
integration algorithms with matching upper and lower bounds. He studied an important class of
integrands - functions which belong to  H\"older spaces. His work is closely related to
information-based complexity - a frame in which the complexity of numerical 
problems is studied (in the classical setting). 

Mainly due to efforts within this theory, by now for many important problems of 
numerical analysis matching upper and lower complexity
bounds (or in other words, optimal convergence rates) are known for both the 
classical deterministic and randomized setting. It is a challenging task to study these
problems in the setting of quantum computation. Once such results are obtained, one can 
compare them to the deterministic and randomized classical ones to understand the 
possible speedups by quantum algorithms. 
Novak (2001) did the  first step toward this, and the present
paper as well as related work, Heinrich and Novak (2001a,b) and Heinrich (2001),
go further along this line. 

In the present paper we study quantum summation of sequences satisfying $p$-summability
conditions. These classes are larger than that of uniformly bounded sequences
(precise definitions are given in section 3) and cannot be handled by the previous 
algorithms. But the solution of this problem is needed for the understanding of
quantum integration in various function spaces (different from H\"older classes) 
characterized by $p$-integrability conditions, such as the Lebesgue spaces $L_p([0,1]^d)$,
studied here in section 5, and the Sobolev spaces analyzed in Heinrich (2001).
In the present paper we therefore develop quantum algorithms for computing the sum of
such sequences. We also prove lower bounds which are, in many cases, matching with
the obtained upper bounds, showing the optimality of the algorithms. (The picture 
is completed in Heinrich and Novak 2001b, where the case is settled which is left open here.)
These results enable us to completely determine (in one case, up to  a logarithmic factor) 
the optimal order of convergence of quantum integration
in Lebesgue spaces $L_p([0,1]^d)$. 

Comparing the result both for summation and integration
with the randomized classical setting, we observe a considerable gain by quantum computing --
the quantum speed of convergence equals the 
square of the randomized classical  one. The gain over deterministic classical algorithms 
can even be exponential (see the details in sections 5 and 6).

To put the problem formulations and the results on a firm mathematical
basis it was necessary to extend the  usual model of quantum computation 
(we follow Beals, Buhrman, Cleve, and Mosca, 1998) to the setting of numerical problems, 
to the fields of
real or complex numbers, normed spaces of functions etc. This extension was widely inspired
by the approach of information-based complexity theory to numerical problems 
in the classical settings and can be viewed, in fact, as a quantum setting of this theory.

The paper is organized as follows. The general approach is presented in section 2. Upper
bounds for summation of $p$-summable sequences and respective algorithms are contained 
in section 3. General results
concerning lower bounds as well as their application to summation are given in section 4.
Section 5 is devoted to the application of the previous results to integration
of functions from the Lebesgue spaces $L_p([0,1]^d)$. Finally, in section 6 we give 
comparisons
to results in the classical deterministic and randomized settings and comment on 
some further related issues.

For background reading in quantum computing we refer to the surveys 
Ekert, Hayden, and Inamori (2000), Shor (2000), and the monographs Pittenger (1999), 
Gruska (1999) and Nielsen and Chuang (2000). For notions and results in in\-for\-ma\-tion-based
complexity theory see the monographs  Traub, Wasilkowski, and 
Wo\'z\-nia\-kowski (1988), Novak (1988), and the survey of the randomized setting 
Heinrich (1993).

\section{A General Framework for Numerical Quantum Algorithms}

We are given nonempty sets $D$, $K$, a nonempty set $F$ of functions on $D$ with values
in $K$ and a function $S$ from $F$ to a normed space $G$. By a normed space we
always mean a normed linear space over $\K$, where $\K$ is either $\R$ or $\C$, 
the field of real or complex numbers.  We seek to compute
(approximately) $S(f)$ for $f\in F$, where $f$ can only be accessed through its
values (that is, we assume that $f$ is given as a black box -- given $t\in D$,
this black box returns $f(t)\in K$).

This general framework includes, on one hand, the binary case, where
$D=\{0,\dots,N-1\}$, $K=\{0,1\}$, $F$ consists of all Boolean functions,
i.\,e.\, all functions from $D$ to $K$, and $S$ maps $F$ to $G=\R$ (which contains
$\{0,1\}$).  On the other hand, in numerical problems, $D$ is usually some subset of
$\R^d$, $K=\K$, $F$ is usually a subset of a normed linear space of
functions (or tuples of functions) from $D$ to $\K$, and $S$ is a mapping 
(also called the solution operator) from $F$ to $G$, where $G$ is either $\K$
or a normed space of functions.

We want to study algorithms and complexity of solving these problems on a
quantum computer. For this purpose, we adopt standard notation of quantum 
computing. Let $H_1$ be the 
two-dimensional complex Hilbert space $\C^2$, $\{e_0,e_1\}$ its unit vector
basis, let 
$$
 H_m=H_1\otimes\dots\otimes H_1 
$$
be the Hilbertian tensor product of $m$ copies of $H_1$. We use the standard
identifications such as writing $e_i$ or $\lg i \rs$ for $
 e_{j_0}\otimes\dots\otimes e_{j_{m-1}} 
$, where $i=\sum_{k=0}^{m-1}j_k2^{m-1-k}$ is the binary expansion of $i$.
When identifying $H_m$ with $H_{m_1}\otimes\dots\otimes H_{m_\ell}$, where 
$\sum_{k=1}^\ell m_j = m$, we also identify 
$e_i$ with the respective $e_{i_1}\otimes\dots\otimes e_{i_\ell}$ and $\lg i_1\rs \dots
\lg i_\ell\rs$, and finally also $i$ itself with $(i_1,\dots,i_\ell)$ in the respective way. 
For convenience we use the following notation:
$$\Z[0,N) := \{0,\dots,N-1\}$$
for $N\in\N$ (as usual, we let $\N=1,2,\dots, \N_0=\N\cup\{0\})$.
Let $\mathcal{C}_m = \{\lg i\rs:\, i\in\Z[0,2^m)\}$ be the set of basis vectors of $H_m$, 
also called classical states, or basis states, and let $\mathcal{U}(H_m)$ denote the 
set of unitary
operators on $H_m$.

First we introduce the notion of a quantum query (in our setting of $D,K,F,G$ and 
$S$). A quantum query  on $F$ is given by a tuple
\begin{equation}
\label{J1}
Q=(m,m',m'',Z,\tau,\beta),
\end{equation}
where $m,m',m''\in \N, m'+m''\le m, Z\subseteq \Z[0,2^{m'})$ is a nonempty 
subset, and
$$\tau:Z\to D$$
$$\beta:K\to\Z[0,2^{m''})$$
are arbitrary mappings. The meaning of these components will be explained below.
Such a tuple $Q$ defines a query mapping (we use the same symbol $Q$) 
$$Q:F\to\mathcal{U}(H_m)$$
$$f\to Q_f$$ 
as follows: Let any $h\in\mathcal{C}_m$ be represented as 
$h=\lg i\rs\lg x\rs\lg y\rs$ with $\lg i\rs\in\mathcal{C}_{m'},
\lg x\rs\in\mathcal{C}_{m''},\lg y\rs\in\mathcal{C}_{m-m'-m''}$ 
(if $m=m'+m''$, we drop the last component). Then $Q_f$ is the unitary operator
defined uniquely by its action on $\mathcal{C}_m$:
\begin{equation}
\label{B1} 
Q_f\lg i\rs\lg x\rs\lg y\rs=
\left\{\begin{array}{ll}
\lg i\rs\lg x\oplus\beta(f(\tau(i)))\rs\lg y\rs &\quad \mbox {if} \quad i\in Z\\
\lg i\rs\lg x\rs\lg y\rs & \quad\mbox{otherwise,} 
 \end{array}
\right. 
\end{equation}
where here and in the sequel $\oplus$ always means addition modulo the 
respective power of 2, here modulo $2^{m''}$. Let $m(Q)$ denote the first component 
of $Q$, that is, the total number of qubits. If $m(Q)=m$, we
also say that $Q$ is an $m$-qubit quantum query. 

 This notion contains the binary 
black box query typically used in quantum computation (see, e.g. Beals, Buhrman, 
Cleve, and Mosca, 1998)
as a particular case: Such a binary query associates to a $\{0,1\}$-valued function
$\kappa$ defined on $\Z[0,2^{m'})$ the operator $Q_\kappa$ which maps 
$\lg i\rs\lg x\rs\lg y\rs$ to $\lg i\rs\lg x\oplus\kappa(i)\rs\lg y\rs $,
where $i\in\mathcal{C}_{m'}$ and $x\in\mathcal{C}_1$.
In our situation we have to deal with two more general domains: $D$ and $K$.
The mapping $\tau : i \to \tau(i)\in D$ describes the (chosen by the algorithm designer)
correspondence of binary strings with certain elements of the domain of definition of 
functions from $F$. Since at request 
$\tau(i)$ the black box returns $f(\tau(i))$, which is an element of $K$, we
need a second mapping $\beta$, which maps ("codes") elements from $K$ into binary
strings. (This is also chosen by the algorithm designer.) As usual, the 
untouched part $\lg y \rs$ stands for "working bits". 

Note that, by the definition,
 a quantum query on $F$
is also a quantum query on any other nonempty subset $F_1\subseteq \mathcal{F}(D,K)$, and in
particular, on $\mathcal{F}(D,K)$ itself. Here $\mathcal{F}(D,K)$ denotes the set of
all functions from $D$ to $K$. Indeed, the mapping $Q_f$ is defined for each
$f\in \mathcal{F}(D,K)$.

Next we define quantum algorithms in the general framework of $D$, $K$, $F$, $G$ and $S$.
It will be convenient for us to introduce algorithms with multiple measurements.
We show later in this section how they can be simulated by algorithms with one 
measurement. Let us first describe informally
what we mean by a quantum algorithm with $k$ measurements: Such an algorithm
starts with a fixed basis state $b_0$ and applies in an alternating way unitary 
transformations (not depending on $f$) and a certain query, associated to
the algorithm. After a fixed
number of steps the resulting state is measured, which gives a (random) basis 
state $\xi_0$. This state is memorized and then transformed (e.g.\ by a classical
computer) into a new basis state $b_1$. This is the starting state to which the
next sequence of quantum operations is applied 
(with possibly another query and number of qubits). The resulting state is again measured, 
which gives the (random) basis state $\xi_1$. This state is memorized, and $b_2$
is computed from $\xi_0$ and $\xi_1$, and so on. After $k$ such cycles, we obtained
$\xi_0,\dots,\xi_{k-1}$. Then finally an element of $G$ -- the output of the algorithm --
is computed (e.g.\ again on a classical computer) from the results of all measurements:
$\varphi(\xi_0,\dots,\xi_{k-1})$.

Now we formalize this:
A quantum algorithm  on $F$  with no measurement is a tuple
\begin{equation*}
A=(Q,(U_j)_{j=0}^n),
\end{equation*}
where $Q$ is a quantum query on $F$, $n\in\N_0$  and
$U_{j}\in \mathcal{U}(H_m)\,(j=0,\dots,n)$, with $m=m(Q)$ (in the case $n=0$,
no query $Q$ is needed).  Given such an $A$ and $f\in F$,
we let $A_f\in \mathcal{U}(H_m)$ be defined as
\begin{equation}
\label{B1a}
A_f = U_n Q_f U_{n-1}\dots U_1 Q_f U_0.
\end{equation}
We denote by $n_q(A):=n$ the number of queries and by $m(A)=m=m(Q)$ the 
number of qubits used by $A$. We also introduce the following notation.
 Let
$A_f(x,y)$ for $x,y\in\Z[0,2^m)$ be given by 
\begin{equation}
\label{14A2}
A_f\lg y\rs=\sum_{x\in\Z[0,2^m)}A_f(x,y)\lg x\rs.
\end{equation}
Hence $(A_f(x,y))_{x,y}$ is the matrix of the transformation $A_f$ in the 
canonical basis $\mathcal{C}_{m}$.

 A quantum algorithm on $F$ with output 
in $G$ (or shortly, from $F$ to $G$) with $k$ measurements  is a tuple
$$
A=((A_\ell)_{\ell=0}^{k-1},(b_\ell)_{\ell=0}^{k-1},\varphi),
$$ 
where $k\in\N,$ and $A_\ell\,(\ell=0,\dots,k-1)$ are quantum algorithms
on $F$ with no measurements. To explain the other components, 
set $m_\ell=m(A_\ell)$. Then 
$$
b_0\in\Z[0,2^{m_0}),
$$
for $1\le \ell \le k-1,\,b_\ell$ is a function
$$
b_\ell:\prod_{i=0}^{\ell-1}\Z[0,2^{m_i}) \to \Z[0,2^{m_\ell}),
$$
and $\varphi$ is a function with values in $G$
$$
\varphi:\prod_{\ell=0}^{k-1}\Z[0,2^{m_\ell}) \to G.
$$
We also say that $A$ is a quantum algorithm with measurement(s), or 
just a quantum algorithm.

Let $\mathcal{P}_0(G)$ denote the set of all probability measures on $G$ whose
support is a finite set.
The output of $A$ at input $f\in F$ will be an element $A(f)\in\mathcal{P}_0(G)$ 
(we use the same symbol $A$ for the mapping $A:F\to \mathcal{P}_0(G)$). 
We define $A(f)$ via a sequence of random variables 
$(\xi_{\ell,f})_{\ell=0}^{k-1}$ (we assume that all random variables are defined over
a fixed -- suitably large -- probability space $(\Omega,\Sigma,\Prm)$).
So let $f\in F$ be fixed. Now let $\xi_{\ell,f}$ be such that
\begin{equation}
\label{14A4}
\Prm\{\xi_{0,f}=x\}=|A_{0,f}(x,b_0)|^2
\end{equation}
and, for $1\le \ell\le k-1$,
\begin{equation}
\label{14A3}
\Prm\{\xi_{\ell,f}=x\,|\,\xi_{0,f}=x_0,\dots,\xi_{\ell-1,f}=x_{\ell-1}\}=
|A_{\ell,f}(x,b_\ell(x_0,\dots,x_{\ell-1}))|^2.
\end{equation} 
Clearly, this defines the distribution of $(\xi_{\ell,f})_{\ell=0}^{k-1}$ uniquely.
Let us define for $x_0\in\Z[0,2^{m_0}),\dots, x_{k-1}\in\Z[0,2^{m_{k-1}})$
\begin{eqnarray}   
p_{A,f}(x_0,\dots, x_{k-1})&=&
|A_{0,f}(x_0,b_0)|^2 |A_{1,f}(x_1,b_1(x_0))|^2\dots\nonumber\\
&&\dots |A_{k-1,f}(x_{k-1},b_{k-1}(x_0,\dots,x_{k-2}))|^2.\label{M1}
\end{eqnarray}
It follows from (\ref{14A4}) and (\ref{14A3}) that
\begin{equation}
\label{14A5}
\Prm\{\xi_{0,f}=x_0,\dots,\xi_{k-1,f}=x_{k-1}\}= p_{A,f}(x_0,\dots, x_{k-1}).
\end{equation}
Finally we define the output of $A$ at input $f$ as
$$
A(f)=\mbox{dist}(\varphi(\xi_{0,f},\dots,\xi_{k-1,f})),
$$
the distribution of $\varphi(\xi_{0,f},\dots,\xi_{k-1,f})$. This random variable 
takes only finitely
many values in $G$, hence the support of $A(f)$ is finite (and no measurability problems 
related to the target space $G$ will arise).
It follows from (\ref{14A5}) that
for any subset $C\subseteq G$
\begin{equation}
\label{M3}
A(f)\{C\}=\sum_{\phi(x_0,\dots,x_{k-1})\in C}p_{A,f}(x_0,\dots, x_{k-1}).
\end{equation}
We note that, analogously to quantum queries, a quantum algorithm
on $F$ is automatically also a quantum algorithm on any nonempty
$F_1\subseteq \mathcal{F}(D,K)$.

The number 
$n_q(A):=\sum_{\ell=0}^{k-1} n_q(A_\ell)$
is called the number of queries used by $A$. This is the crucial quantity for 
the purposes of our query
complexity analysis. (In section 6 we give some comments on the  
cost in the bit-model.)

Let $0\le\theta< 1$. For an algorithm $A$ as above  we 
define the (probabilistic) error at $f\in F$ as follows. 
Let 
$\zeta$ be a random variable with distribution $A(f)$. Then 
\begin{equation*}
e(S,A,f,\theta)=\inf\left\{\varepsilon\,\,|\,\,\Prm\{\|S(f)-\zeta\|>\varepsilon\}\le\theta
\right\}
\end{equation*}
(note that this infimum is always attained).
Hence $e(S,A,f,\theta)\le \varepsilon$ iff the algorithm $A$ computes $S(f)$  with 
error at most $\varepsilon$ and probability at least $1-\theta$. 
We put 
$$
e(S,A,F,\theta)=\sup_{f\in F} e(S,A,f,\theta) 
$$
(we allow the value $+\infty$ for this quantity). Furthermore, we set
$$
e(S,A,f)=e(S,A,f,1/4)
$$
and similarly,
$$
e(S,A,F)=e(S,A,F,1/4).
$$
The central quantity of our study is the 
$n$-th minimal (query) error, defined for $n\in\N_0$ by 
$$
e_n^q(S,F)=\inf\{e(S,A,F)\,\,|\,\,A\,\,\mbox{is any quantum algorithm with}\,\, n_q(A)\le n\},
$$
that is, the smallest error which can be reached using at most $n$ queries.
The query complexity is defined for $\varepsilon > 0$ by 
\begin{eqnarray*}
\lefteqn{\mbox{comp}_\varepsilon^q(S,F)=}\\
&&\min\{n_q(A)\,\,|\,\, A\,\,\mbox{is any quantum 
algorithm with}\,\, e(S,A,F) \le \varepsilon\}
\end{eqnarray*}
(we put $\mbox{comp}_\varepsilon^q(S,F)=+\infty$ if there is no such algorithm).
It is easily checked that these functions are inverse to each other in the 
following sense: For all $n\in \N_0$ and $\varepsilon > 0$,
$e_n^q(S,F)\le \varepsilon$ if and only if
$\mbox{comp}_{\varepsilon_1}^q(S,F)\le n$ for all $\varepsilon_1 > \varepsilon$.
Hence it suffices to determine one of them. We shall usually choose the first one.

Our first general result shows the tight relation between algorithms with
several measurements and (the conceptually simpler)
algorithms with one measurement. It states that an algorithm with several measurements
can always be represented equivalently by an algorithm with one measurement and
twice the number of queries  
(at the expense of an increased number of qubits).
\begin{lemma}
\label{lem:1}
For each quantum algorithm $A$ from $F$ to $G$ with $k$ measurements there is a
quantum algorithm $\widetilde{A}$ from $F$ to $G$ with one measurement such that 
$n_q(\widetilde{A})=2n_q(A)$ and
$$
\widetilde{A}(f)=A(f)
$$
for all $f\in F$.
\end{lemma}
\begin{proof}
By '$\ell$-th quantum cycle' we mean the quantum operations in the original algorithm
before the first measurement if $\ell=0$, and between the $\ell$-th and
the $\ell+1$-st measurement if $1\le\ell\le k-1$.
The idea of the proof is easy: We simulate the $k$ queries by one
query and instead of intermediate measurements we 'store' the results of the 
cycles in different 
components until the final measurement (a pseudo-code is given below). 
Let us now formalize this and check 
that the corresponding probabilities coincide.
 Let the original algorithm be given by
$$
A=((A_\ell)_{\ell=0}^{k-1},(b_\ell)_{\ell=0}^{k-1},\varphi),
$$ 
where 
$$
A_\ell=(Q_\ell,(U_{\ell j})_{j=0}^{n_\ell}),
$$
and
$$
Q_\ell=(m_\ell,m'_\ell,m''_\ell,Z_\ell,\tau_\ell,\beta_\ell).
$$
By adding, if necessary, qubits, which are set to zero and remain so during the whole
$\ell$-th cycle we may assume without loss of generality that $m'_\ell\equiv m'$.
Let\footnote{Throughout this paper $\log$ stands for $\log_2$.} 
$k_0 = \lceil \log k\rceil$, define $\widetilde{m}'=m'+k_0$, and
$\widetilde{Z}\subset\Z[0,2^{\widetilde{m}'})$ by
$$
\widetilde{Z}=\{(\ell,i)\, |\, 0\le\ell\le k-1,\,i\in Z_\ell\}.
$$
 Now we define 
$$
\wt{\tau}:\widetilde{Z} \to D
$$
$$
\wt{\tau}(\ell,i)=\tau_\ell(i)\quad \mbox{for}\quad (\ell,i)\in \widetilde{Z}.
$$
 Moreover, we set 
$$
\widetilde{m}''=\sum_{\ell=0}^{k-1} m''_\ell,
$$ 
$$
\wt{\beta}: K\to \Z[0,2^{\widetilde{m}''})
$$
$$
\wt{\beta}(s)=(\beta_0(s),\dots,\beta_{k-1}(s))\quad \mbox{for} \quad s\in K. 
$$
$$
\widetilde{m}=k_0+\widetilde{m}''+\sum_{\ell=0}^{k-1} m_\ell
$$
$$
\widetilde{Q}=(\widetilde{m},\widetilde{m}',\widetilde{m}'',\widetilde{Z},
\wt{\tau},\wt{\beta}).
$$
Let us fix the following notation: Consider the splitting 
$$
 H_{\widetilde{m}}=H_{k_0}\otimes H_{\widetilde{m}''}\otimes H_{m_0}
 \otimes\dots\otimes H_{m_{k-1}}. 
$$
The representation of a basis state
$$
\lg i\rs\lg u\rs\lg x_0\rs\dots\lg x_{k-1}\rs
$$
refers to this splitting. We also need refined splittings. We represent
$$
H_{\widetilde{m}''}=H_{m''_0}\otimes\dots\otimes H_{m''_{k-1}}, 
$$
and
$$
\lg u\rs=\lg u_0\rs\dots\lg u_{k-1}\rs 
$$
corresponds to that splitting. Similarly,
$$
H_{m_\ell}=H_{m'}\otimes H_{m''_\ell}\otimes H_{m-m'-m''_\ell}
$$
with the respective
$$
\lg x_{\ell}\rs=\lg i_{\ell} \rs\lg y_{\ell} \rs\lg z_{\ell} \rs.
$$
Next we define the following unitary operators on $H_{\widetilde{m}}$
by their action on the basis states:
$$
J \lg i\rs\lg u\rs\lg x_0\rs\dots\lg x_{k-1}\rs
=\lg i\rs\lg \ominus u\rs\lg x_0\rs\dots\lg x_{k-1}\rs,
$$
where $\ominus$ means subtraction modulo $2^{\widetilde{m}''}$ and $\ominus u$ stands for
$0\ominus u$,
$$
C \lg i\rs\lg u\rs\lg x_0\rs\dots\lg x_{k-1}\rs
=\lg i\oplus 1\rs\lg u\rs\lg x_0\rs\dots\lg x_{k-1}\rs,
$$
for $\ell=0,\dots,k-1$, $j=0,\dots,n_\ell$,
\begin{eqnarray*}
& & T_\ell\lg i\rs\lg u_0\rs\dots\lg u_\ell\rs\dots\lg u_{k-1}\rs
\lg x_0\rs\dots\lg i_{\ell} \rs\lg y_{\ell} \rs\lg z_{\ell} \rs     
\dots\lg x_{k-1}\rs 
\\
& &=\lg i\rs\lg u_0\rs\dots\lg u_\ell\rs\dots\lg u_{k-1}\rs
\lg x_0\rs\dots\lg i_{\ell} \rs\lg y_{\ell}\oplus u_\ell \rs\lg z_{\ell} \rs     
\dots\lg x_{k-1}\rs 
\end{eqnarray*}
$$
\widetilde{U}_{\ell j}\lg i\rs\lg u\rs
\lg x_0\rs\dots\lg x_\ell\rs\dots\lg x_{k-1}\rs
=\lg i\rs\lg u\rs
\lg x_0\rs\dots(U_{\ell j}\lg x_\ell\rs)\dots\lg x_{k-1}\rs,
$$
\begin{eqnarray*}
& & P_\ell \lg i\rs\lg u\rs\lg x_0\rs\dots
\lg i_{\ell} \rs\lg y_{\ell} \rs\lg z_{\ell} \rs     
\dots\lg x_{k-1}\rs 
\\
& & =\lg i\rs\lg i_{\ell} \rs\lg u\rs
\lg x_0\rs\dots\lg y_{\ell}\rs\lg z_{\ell} \rs     
\dots\lg x_{k-1}\rs,\\
\end{eqnarray*}
and finally, for $\ell=1,\dots,k-1$, 
\begin{eqnarray*}
& & B_\ell\lg i\rs\lg u\rs
\lg x_0\rs\dots\lg x_{\ell-1}\rs\lg x_\ell\rs\dots\lg x_{k-1}\rs\\
& & =\lg i\rs\lg u_0\rs\lg x_0\rs\dots\lg x_{\ell-1}\rs\lg x_\ell
\oplus b_\ell(x_0,\dots,x_{\ell-1})\rs\dots\lg x_{k-1}\rs.
\end{eqnarray*}
Now we present the simulation of the queries $Q_{\ell, f}$ by $\widetilde{Q}_f$:
Let $0\le \ell\le k-1$. It is readily checked, that if we apply the operator 
$P_{\ell}^{-1}\widetilde{Q}_f P_{\ell}$ to the state 
$$
\lg \ell\rs\lg 0\rs\lg x_0\rs\dots\lg x_{k-1}\rs,
$$
we get
\begin{eqnarray*}
&&\lg \ell\rs\lg \wt{\beta}(f(\wt{\tau}(\ell,i_\ell)))\rs
\lg x_0\rs\dots\dots\lg x_{k-1}\rs
\\
&=&\lg \ell\rs\lg \beta_0(f(\tau_\ell(i_\ell)))\rs\dots
\lg \beta_{k-1}(f(\tau_\ell(i_\ell)))\rs
\lg x_0\rs\dots\dots\lg x_{k-1}\rs,
\end{eqnarray*}
provided $i_\ell\in Z_\ell$. Applying then $T_\ell$ to this state gives
  
\begin{eqnarray*}
&&\lg \ell\rs\lg \wt{\beta}(f(\wt{\tau}(\ell,i_\ell)))\rs
\lg x_0\rs\dots\lg i_{\ell} \rs\lg y_{\ell}\oplus\beta_\ell(f(\tau_\ell(i_\ell)))\rs
\lg z_{\ell} \rs\dots\lg x_{k-1}\rs
\\
&=&\lg \ell\rs\lg \wt{\beta}(f(\wt{\tau}(\ell,i_\ell)))\rs
\lg x_0\rs\dots (Q_{\ell ,f}\lg x_\ell\rs)\dots\lg x_{k-1}\rs.
\end{eqnarray*}
Next $J$ is applied which yields
$$
\lg \ell\rs\lg \ominus\wt{\beta}(f(\wt{\tau}(\ell,i_\ell)))\rs
\lg x_0\rs\dots (Q_{\ell,f}\lg x_\ell\rs)\dots\lg x_{k-1}\rs,
$$
and finally the application of $P_{\ell}^{-1}\widetilde{Q}_f P_{\ell}$ produces
$$
\lg \ell\rs\lg 0\rs\lg x_0\rs\dots (Q_{\ell ,f}\lg x_\ell\rs)\dots\lg x_{k-1}\rs.
$$
If $i_\ell\not\in Z_\ell$, this also holds, which is checked in the same way.
Hence we showed that 
\begin{eqnarray}
\lefteqn{P_{\ell}^{-1}\widetilde{Q}_f P_{\ell}J T_\ell P_{\ell}^{-1}\widetilde{Q}_f P_{\ell}
\lg \ell\rs\lg 0\rs\lg x_0\rs\dots \lg x_\ell\rs\dots\lg x_{k-1}\rs}\hspace{3cm}\nonumber\\ 
&=&
\lg \ell\rs\lg 0\rs\lg x_0\rs\dots (Q_{\ell ,f}\lg x_\ell\rs)\dots\lg x_{k-1}\rs.
\end{eqnarray}
The new algorithm can now be described as follows:
\begin{tabbing}
\hspace{1cm}\=\hspace{1cm}\=\hspace{1cm}\= \kill
\>initialize $\lg 0\rs\lg 0\rs\lg b_0\rs\lg 0\rs\dots\lg 0\rs$\\
\> for $\ell=0,\dots,k-1$ do\\
\>\> apply $\widetilde{U}_{\ell,0}$
\quad (beginning of $\ell$-th cycle of original algorithm)\\
\>\>for $j=1,\dots,n_\ell$\\
\>\>\>apply $P_{\ell}^{-1}\widetilde{Q}_f P_{\ell}J T_\ell P_{\ell}^{-1}\widetilde{Q}_f P_{\ell}$\\
\>\>\>apply $\widetilde{U}_{\ell j}$\quad (end of $\ell$-th cycle of original algorithm)\\
\>\>if $\ell\neq k-1$\\ 
\>\>\>apply $B_{\ell+1}$ \quad (computing $b_{\ell+1}$ as initial state of next\\ 
\>\>\>cycle)\\
\>\>\>apply $C$\quad (increasing the counter by one)\\
\>measure all qubits corresponding to the components\\ 
\>$H_{m_0},\dots,H_{m_{k-1}}$ (let $\lg x_0\rs\dots\lg x_{k-1}\rs$ be the result)\\
\>compute $\varphi(x_0,\dots,x_{k-1})$.
\end{tabbing}
The starting passage through the outer loop ($\ell=0$) acts as follows:
$$
\lg 0\rs\lg 0\rs\lg b_0\rs\lg 0\rs ^{(k-1)} \to 
\lg 1\rs\lg 0\rs\left( \sum_{x_0}A_{0,f}(x_0,b_0) 
\lg x_0\rs\lg b_1(x_0)\rs\right) \lg 0\rs ^{(k-2)}.
$$
The passage with index $ \ell $, $1\le \ell \le k-2$,  maps each basis state
of the form
$$
\lg \ell\rs\lg 0\rs\lg x_0\rs\dots\lg x_{\ell -1}\rs\lg y\rs\lg 0\rs ^{(k-\ell-1)}
$$
into
$$
\lg \ell+1\rs\lg 0\rs\lg x_0\rs\dots\lg x_{\ell-1}\rs
\left( \sum_{x_\ell}A_{\ell ,f}(x_\ell,y) 
\lg x_\ell\rs\lg b_{\ell+1}(x_0,\dots,x_\ell)\rs\right) \lg 0\rs ^{(k-\ell-2)}.
$$
Finally, the last passage  ($ \ell=k-1$) acts as follows:
\begin{eqnarray*}
\lefteqn{
\lg k-1\rs\lg 0\rs\lg x_0\rs\dots\lg x_{k-2}\rs\lg y\rs \to
}\\
&&\lg k-1\rs\lg 0\rs\lg x_0\rs\dots\lg x_{k-2}\rs
\left( \sum_{x_{k-1}}A_{k-1,f}(x_{k-1},y) \lg x_{k-1}\rs\right) .
\end{eqnarray*}
From this it follows that the overall result of the algorithm before measurement
is the state
\begin{eqnarray*}
\lefteqn{
\sum_{x_0,\dots,x_{k-1}}A_{0,f}(x_0,b_0)A_{1,f}(x_1,b_1(x_0))
\dots
}\\
&&\dots A_{k-1,f}(x_{k-1},b_{k-1}(x_0,\dots,x_{k-2}))
\lg k-1\rs\lg 0\rs\lg x_0\rs\dots\lg x_{k-1}\rs .
\end{eqnarray*}
The probability of measuring $\lg x_0\rs\dots\lg x_{k-1}\rs$ is thus
$$
|A_{0,f}(x_0,b_0)|^2 |A_{1,f}(x_1,b_1(x_0))|^2
\dots |A_{k-1,f}(x_{k-1},b_{k-1}(x_0,\dots,x_{k-2}))|^2,
$$
which equals
$$
\Prm\{ \xi_{0,f}=x_0,\dots,\xi_{k-1,f}=x_{k-1}\},
$$
by (\ref{M1}) and (\ref{14A5}). This proves the lemma.
\end{proof}
We will sometimes write that we repeat a quantum algorithm a number of times,
or, more generally, that we apply to $f\in F$ a finite sequence of algorithms 
$A_i$ from $F$ to $G_i$  ($i=0,\dots,M-1$) and combine the results
by the help of a classical computation. 
Let
$$
\psi:G_0\times\dots\times G_{M-1} \to G
$$
be any mapping. 
 Using our notion of a 
quantum algorithm with measurements, a formal representation of the composed 
algorithm $A$, which we write symbolically as
\begin{equation}
\label{A0}
A=\psi (A_0,\dots,A_{M-1}),
\end{equation}
can easily be given as follows: Let 
$$
A_i=((A_{i,\ell})_{\ell=0}^{k_i-1},(b_{i,\ell})_{\ell=0}^{k_i-1},\varphi_i),
$$
put $k=\sum_{i=0}^{M-1} k_i$, let the set
$$
\Upsilon=\{ (i,\ell)\,\,|\,\, i=0,\dots,M-1,\, \ell=0,\dots,k_i-1\}
$$
be equipped with the lexicographical order, and let
$$
\phi=\psi(\phi_0,\dots,\phi_{M-1}).
$$
Then we define
$$
\psi (A_0,\dots,A_{M-1})=((A_\upsilon)_{\upsilon\in\Upsilon},
(b_\upsilon)_{\upsilon\in\Upsilon},\phi).
$$
The next lemma gives some further description of the composition and 
is readily checked using the definition of a quantum algorithm.
We need the following notation: For probability measures 
$\mu_0,\dots,\mu_{M-1}\in \mathcal{P}_0(G)$ let 
$\psi(\mu_0,\dots,\mu_{M-1})\in \mathcal{P}_0(G)$ be the measure induced by 
$\mu_0\times\dots\times\mu_{M-1}$
via $\psi$ on $G$, that is, for $C\subseteq G$,
$$
\psi(\mu_0,\dots,\mu_{M-1})(C)=(\mu_0\times\dots\times\mu_{M-1})(\psi^{-1}(C)).
$$
\begin{lemma}
\label{lem:2a}
For each $f\in F$,
$$ 
\psi (A_0,\dots,A_{M-1})(f)=\psi (A_0(f),\dots, A_{M-1}(f)),
$$
or stated equivalently, if  $(\zeta_i)_{i=0}^{M-1}$ are independent random 
variables with distribution $A_i(f)$ respectively, then  
$$
\psi (A_0,\dots,A_{M-1})(f)={\rm dist}(\psi(\zeta_0,\dots,\zeta_{M-1})).
$$
Moreover, 
$$
n_q(\psi (A_0,\dots, A_{M-1}))=\sum_{i=0}^{M-1}n_q(A_i).
$$
\end{lemma}
The next lemma concerns the special
case of repeating an algorithm. It describes a standard technique of boosting
the success probability. For completeness, we include the short proof. 
Let $G=\R$, $M\in \N$ and denote by 
$\psi_0:\R^M\to \R$ the mapping given by the median, that is, 
$\psi_0(a_0,\dots,a_{M-1})$ is the value of the 
of the $\lceil (M+1)/2\rceil$-th element of the
non-decreasing rearrangement of $(a_i)$. For any algorithm $A$ from $F$ to $\R$ denote
$\psi_0(A^M):= \psi_0(A,\dots,A)$.
\begin{lemma}
\label{lem:2e}
Let $A$ be any quantum algorithm and $S$ be any mapping from $F$ to $\R$. 
Then for each $f\in F$,
$$
e(S,\psi_0(A^M),f,e^{-M/8})\le e(S,A,f).
$$
\end{lemma}
\begin{proof}
Fix $f\in F$.
Let $\zeta_0,\dots,\zeta_{M-1}$ be independent random variables with distribution $A(f)$.
Let $\chi_i$ be the indicator function of the set
$\{|S(f)-\zeta_i|>e(S,A,f)\}$.
Then $\Prm\{\chi_i=1\}\le 1/4$. Hoeffding's inequality, see e.g.\ Pollard (1984), p.\ 191,
yields
$$
\Prm\left\{\sum_{i=0}^{M-1}\chi_i\ge M/2\right\}\le
\Prm\left\{\sum_{i=0}^{M-1}(\chi_i-\E\chi_i)\ge M/4\right\}\le e^{-M/8}.
$$
Hence, with probability at least $1-e^{-M/8}$, 
$$
\big|\{i\,\,|\,\,|S(f)-\zeta_i|\le e(S,A,f)\}\big|> M/2,
$$
which implies
$$
\left|S(f)-\psi_0(\zeta_0,\dots,\zeta_{M-1})\right|\le e(S,A,f).
$$
\end{proof}

Another way of building new algorithms from previous ones will also be important for us.
To explain it, let $\emptyset\ne F\subseteq \mathcal{F}(D,K)$ and 
$\emptyset\ne \wt{F}\subseteq \mathcal{F}(\wt{D},\wt{K})$, where $D,\wt{D},K,\wt{K}$ are nonempty sets.
In the construction of a new algorithm $A$ on $F$ we sometimes construct from $f$ a function 
$\wt{f}=\Gamma(f)\in \wt{F}$ to which we want to apply an already developed 
algorithm $\wt{A}$ on $\wt{F}$. 
By definition, the  algorithm $A$ on $F$ can only use queries $Q$ on $F$
itself, while we need to use $\wt{Q}_{\Gamma(f)}$, where $\wt{Q}$ is a query on $\wt{F}$. 
Nevertheless often a solution 
can be found as follows: We simulate $\wt{Q}_{\Gamma(f)}$ either as $Q_f$ with a 
suitable query $Q$  on $F$ or as $B_f$, where  
$B$ is an algorithm without measurement on $F$. The details are given below. 

The first result covers the simple situation where one query 
is just replaced by another.
Let $\eta:\wt{D}\to D$ and $\rho:K\to \wt{K}$
be arbitrary mappings and define $\Gamma:F\to \wt{F}$ by
\begin{equation}
\label{F2}
\Gamma(f) = \rho\circ f\circ\eta.
\end{equation}
\begin{lemma}
\label{lem:2b}
Let $\Gamma$ be a mapping of the form (\ref{F2}). Then 
for each query $\wt{Q}$ on $\wt{F}$ there is a query 
$Q$ on $F$ such that $m(Q)=m(\wt{Q})$ and for all $f\in F$
$$
Q_f=\wt{Q}_{\Gamma(f)}.
$$   
\end{lemma}
\begin{proof} Let 
$$
\wt{Q}=(\wt{m},\wt{m}',\wt{m}'',\wt{Z},\wt{\tau},\wt{\beta}).
$$
Then we define 
$$
Q=(\wt{m},\wt{m}',\wt{m}'',\wt{Z},\tau,\beta),
$$
where $\tau=\eta\circ\wt{\tau}$ and $\beta=\wt{\beta}\circ\rho$.
Now the lemma follows directly from the query definition. 
\end{proof}
The second result in this direction is slightly more technical. We assume that we are
given a mapping $\Gamma:F\to \wt{F}$ of the following type: There are an 
$m^*\in\N$ and mappings
\begin{eqnarray*}
\eta&:& \wt{D}\to D \\
\beta&:& K\to\Z\big[0,2^{m^*}\big)\\
\rho&:& \wt{D}\times\Z\big[0,2^{m^*}\big) \to \wt{K}
\end{eqnarray*}
such that for $f\in F$ and $s\in \wt{D}$
\begin{equation}
\label{F1}
\Gamma (f)(s)=\rho(s,\beta\circ f\circ\eta(s)).
\end{equation}
\begin{lemma}
\label{lem:2c}
Let  $\wt{Q}$ be a quantum query on $\wt{F}$ and let $\Gamma$
 be a mapping  of the above form (\ref{F1}). Then
there is a quantum algorithm without measurement $B$ on $F$ such that 
$n_q(B)= 2$, $m(B)=m(\wt{Q})+m^*$
and for all $f\in F$, $x\in \Z\big[0,2^{m(\wt{Q})}\big)$,
$$
B_f\lg x\rs\lg 0\rs_{m^*}=(\wt{Q}_{\Gamma(f)}\lg x\rs)\lg 0\rs_{m^*},
$$
where $\lg 0\rs_{m^*}$ stands for the zero state
in $\Z[0,2^{m^*})$.
\end{lemma}
\begin{proof} Let 
$$
\wt{Q}=(\wt{m},\wt{m}',\wt{m}'',\wt{Z},\wt{\tau},\wt{\beta}),
$$
and put
\begin{eqnarray*}
&m=\wt{m}+m^*,\quad m'=\wt{m}', \quad
m''=m^*,\\
&Z=\wt{Z},\quad
\tau=\eta\circ\wt{\tau}, 
\end{eqnarray*}
let $\beta$ be as above and define
$$
Q=(m,m',m'',Z,\tau,\beta).
$$
We represent 
$$
 H_{m}=H_{\wt{m}'}\otimes H_{\wt{m}''}\otimes H_{\wt{m}-\wt{m}'-\wt{m}''}
 \otimes H_{m^*},
$$
a basis state of which will be written as
$$
\lg i\rs\lg x\rs\lg y\rs\lg z\rs.
$$
Define the permutation operator $P$ by
$$
P\lg i\rs\lg x\rs\lg y\rs\lg z\rs
=\lg i\rs\lg z\rs\lg x\rs\lg y\rs,
$$
the operator of sign inversion
$$
J\lg i\rs\lg z\rs\lg x\rs\lg y\rs
=\lg i\rs\lg \ominus z\rs\lg x\rs\lg y\rs,
$$
and finally
$$
T\lg i\rs\lg z\rs\lg x\rs\lg y\rs
=\lg i\rs\lg z\rs\lg x\oplus\wt{\beta}\circ\rho(\wt{\tau}(i),z)\rs\lg y\rs
$$
if $i\in Z$, and 
$$
T\lg i\rs\lg z\rs\lg x\rs\lg y\rs=\lg i\rs\lg z\rs\lg x\rs\lg y\rs
$$
if $i\not\in Z$.
We define $B$ by setting for $f\in F$,
$$
B_f=P^{-1}Q_f J T Q_f  P.
$$
Let us trace the action of $B_f$ on
$$
\lg i\rs\lg x\rs\lg y\rs\lg 0\rs.
$$
First we assume $i\in Z$. The transformation $Q_fP$ leads to
$$
\lg i\rs
\lg\beta(f(\tau(i)))\rs\lg x\rs\lg y\rs=
\lg i\rs
\lg\beta\circ f\circ\eta\circ\wt{\tau}(i)\rs\lg x\rs\lg y\rs.
$$
 Then the above is mapped by $T$ to
\begin{eqnarray*}   
&&\lg i\rs\lg \beta\circ f\circ\eta\circ\wt{\tau}(i)\rs
\lg x\oplus\wt{\beta}\circ\rho(\wt{\tau}(i),\beta\circ f\circ\eta\circ\wt{\tau}(i))\rs\lg y\rs\\
&=&\lg i\rs\lg\beta(f(\tau(i)))\rs
\lg x\oplus \wt{\beta}(\Gamma (f)(\wt{\tau}(i)))\rs\lg y\rs,
\end{eqnarray*}
and $P^{-1}Q_fJ$ gives
$$
\lg i\rs\lg x\oplus \wt{\beta}(\Gamma (f)(\wt{\tau}(i)))\rs\lg y\rs\lg 0\rs
=\big(\wt{Q}_{\Gamma(f)}\lg i\rs\lg x\rs\lg y\rs\big)\lg 0\rs.
$$
The case $i\not\in Z$ is checked analogously.
\end{proof}
\begin{corollary}
\label{cor:1}
Given a mapping $\Gamma:F\to \wt{F}$ as in  (\ref{F2}) or (\ref{F1}), a normed space
$G$ and a quantum algorithm $\wt{A}$ from $\wt{F}$ to $G$, there is a quantum algorithm 
$A$ from $F$ to $G$ with 
\[
n_q(A)=\left\{\begin{array}{rll}
   n_q(\wt{A})& \mbox{in case of (\ref{F2})}      \\
   2\,n_q(\wt{A})&\mbox{in case of (\ref{F1})}\\
    \end{array}
\right. 
\]
and for all $f\in F$
$$
A(f)=\wt{A}(\Gamma(f)).
$$
Consequently, if $\wt{S}:\wt{F}\to G$ is any mapping and $S=\wt{S}\circ\Gamma$,
then for each $n\in \N_0$
\begin{eqnarray*}
e_n^q(S,F)&\le& e_n^q(\wt{S},\wt{F})\quad \mbox{in case of (\ref{F2}), and}      \\
e_{2n}^q(S,F)&\le& e_n^q(\wt{S},\wt{F})\quad \mbox{in case of (\ref{F1})}.
\end{eqnarray*}
\end {corollary} 
\begin{proof} Let
$$
\wt{A}=((\wt{A}_\ell)_{\ell=0}^{k-1},(\wt{b}_\ell)_{\ell=0}^{k-1},\wt{\phi}),\quad 
\wt{A}_\ell=(\wt{Q}_\ell,(\wt{U}_{\ell,j})_{j=0}^{n_\ell}),
$$
and $\wt{m}_\ell=m(\wt{A}_\ell)$. Then for $f\in F$, $0\le\ell<k$,
$$
\wt{A}_{\ell,\Gamma(f)} = \wt{U}_{\ell,n_\ell} \wt{Q}_{\ell,\Gamma(f)} \wt{U}_{\ell,n_\ell-1}\dots 
\wt{U}_{\ell,1} \wt{Q}_{\ell,\Gamma(f)} \wt{U}_{\ell,0}.
$$
In case of (\ref{F2}) we obtain $A$ by just replacing $\wt{Q}_\ell$ by $Q_\ell$ 
from Lemma \ref{lem:2b}.
It follows from (\ref{M1}) and (\ref{M3}) that
$$
A(f)=\wt{A}(\Gamma(f)).
$$
In case of (\ref{F1}) we replace $\wt{Q}_\ell$ by $B_\ell$ from Lemma \ref{lem:2c},
$\wt{U}_{\ell,j}$ by $U_{\ell,j}=\wt{U}_{\ell,j}\otimes Id_{H_{m^*}}$, where $Id_{H_{m^*}}$
is the identity on $H_{m^*}$, the state
$\lg\wt{b}_0\rs$ by $\lg b_0\rs =\lg \wt{b}_0\rs \lg 0\rs_{m^*}$
and, for $1\le \ell\le k-1$, the mappings 
$$
\wt{b}_\ell:\prod_{i=0}^{\ell-1}\Z[0,2^{\wt{m}_i}) \to \Z[0,2^{\wt{m}_\ell})
$$
by
$$
b_\ell:\prod_{i=0}^{\ell-1}\big(\Z[0,2^{\wt{m}_i})\times\Z[0,2^{m^*})\big) 
\to \Z[0,2^{\wt{m}_\ell})\times\Z[0,2^{m^*}),
$$
defined by
$$
b_\ell((x_0,y_0),\dots,(x_{\ell-1},y_{\ell-1}))
=\big(\wt{b}_\ell(x_0,\dots,x_{\ell-1}),0\big).
$$
Finally, we replace 
$$
\wt{\varphi}:\prod_{\ell=0}^{k-1}\Z[0,2^{\wt{m}_\ell}) \to G
$$
by
$$
\varphi:\prod_{\ell=0}^{k-1}\big(\Z[0,2^{\wt{m}_\ell})\times\Z[0,2^{m^*})\big) \to G,
$$
defined as
$$
\phi((x_0,y_0),\dots,(x_{k-1},y_{k-1}))=\wt{\phi}(x_0,\dots,x_{k-1}).
$$
It follows that 
\[
A_{\ell,f}((x,y),(z,0))=\left\{\begin{array}{lll}
   \wt{A}_{\ell,\Gamma(f)}(x,z)& \mbox{if} \quad y=0  \\
   0& \mbox{otherwise,}    \\
    \end{array}
\right. 
\]
and therefore, by (\ref{M1}), 
\begin{eqnarray*}
\lefteqn{ p_{A,f}((x_0,y_0)\dots,(x_{k-1},y_{k-1}))}\hspace{2cm}\\ 
&=&\left\{\begin{array}{lll}
 p_{\wt{A},\Gamma(f)}(x_0,\dots,x_{k-1})  & \mbox{if} \quad y_0=\dots=y_{k-1}=0   \\
   0& \mbox{otherwise,}    \\
    \end{array}
\right. 
\end{eqnarray*}
which together with (\ref{M3}) yields
$$
A(f)=\wt{A}(\Gamma(f)).
$$
 This proves the first part of the statement. 
The second part is an obvious consequence.
\end{proof}
Finally we state some elementary but useful properties of $e_n^q$.
For $\lambda\in \K$ define  $\lambda S:F\to G$ by 
$(\lambda S)(f)=\lambda S(f)$\, $(f\in F)$. Furthermore, in the case  $K=\K$ we denote
$\lambda F=\{\lambda f\,|\,f\in F\}$.
\begin{lemma}
\label{lem:2d}
Let $S,T:F\to G$ be any mappings, $n\in\N_0$ and assume that $e_n^q(S,F)$ is finite.
Then the following hold:\\
(i)
$$
e_n^q(T,F)\le e_n^q(S,F)+\sup_{f\in F}\|T(f)-S(f)\|.
$$
(ii) For each $\lambda\in\K$
$$
e_n^q(\lambda S,F)=|\lambda|e_n^q(S,F).
$$
(iii) If $K=\K$ and $S$ is a linear operator from $\mathcal{F}(D,K)$ to $G$, then 
for all $\lambda\in\K$
$$
e_n^q(S,\lambda F)=|\lambda|e_n^q(S,F).
$$  
\end{lemma}
\begin{proof}
The first two statements are simple consequences of the definitions. Let us verify the
third one. Let $\wt{F}=\lambda F$, $\Gamma:F\to \wt{F}$ be defined as
$\Gamma(f)=\lambda f$, which is of the form (\ref{F2}). We assume $\lambda \ne 0$, the
case $\lambda=0$ follows trivially from (ii). Since $S$ is linear, we have 
$$
\lambda^{-1} S\circ\Gamma=S,
$$
and hence, by Corollary \ref{cor:1} and statement (ii) above,
$$
e_n^q(S,F)\le |\lambda|^{-1}e_n^q(S,\wt{F})=|\lambda|^{-1}e_n^q(S,\lambda F).
$$
Replacing $F$  by $\lambda F$ and $\lambda$ by $\lambda ^{-1}$, we get
$$
e_n^q(S,\lambda F)\le |\lambda|e_n^q(S,F),
$$
which completes the proof.  
\end{proof}

\section{Quantum Summation}
In this section we study summation of sequences or, what is essentially the same,
the computation of the mean, on a quantum computer. For a fixed $N\in\N$
we set $D=\Z[0,N)$, $K=\R$, $G=\R$, and for  $1\le p\le\infty$ 
let $L_p^N$ denote the space of all functions
$f:D\to \R$, equipped with the norm 
$$
\|f\|_{L_p^N}=\left(\frac{1}{N}\sum_{i=0}^{N-1}|f(i)|^p \right)^{1/p}
$$
if $p<\infty$
and
$$
\|f\|_{L_\infty^N}=\max_{0\le i\le N-1} |f(i)|.
$$
(Note that $L_p^N$ is just the space $L_p(D,\mu)$, 
where $\mu$ is the equidistribution on $D$.) Define $S_N:L_p^N\to\R$ by 
$$
S_N f=\frac{1}{N}\sum_{i=0}^{N-1}f(i). 
$$  We let
$$
F=\mathcal{B}_p^N:= \mathcal{B}(L_p^N)=\{f\in L_p^N \,|\, \|f\|_{L_p^N}\le 1\}
$$
be the unit ball of $L_p^N$. 
 We also define
$$
\mathcal{B}_{\infty,+}^N=\left\{ f:D\to \R\,|\, 0\le f(i)\le 1\,\,\mbox{for all}\,\,i\right\} 
$$
and 
$$
\mathcal{B}_{\infty,0}^N=\left\{ f:D\to \{0,1\} \right\}. 
$$
When we consider $\mathcal{B}_{\infty,0}^N$, we put $K=\{0,1\}$. Clearly,
$$
\mathcal{B}_{\infty,0}^N\subset \mathcal{B}_{\infty,+}^N\subset \mathcal{B}_p^N\subset \mathcal{B}_q^N
$$
whenever $1\le q<p<\infty$. Therefore, we will also consider $S_N$ as acting
on $ \mathcal{B}_{\infty,0}^N$ and $\mathcal{B}_{\infty,+}^N$. 
We use the following standard representations
depending on 
the range of $f\in F$: Given $a,b\in\R,\,a<b$, and $\kappa\in \N$, define 
$\beta_{\kappa,a,b}:\R\to\Z[0,2^{\kappa})$ by
\begin{eqnarray}
\label{N1}
\beta_{\kappa,a,b}(x)=
\left\{\begin{array}{ll}
   2^{\kappa}-1 &\quad\mbox{if}\quad x\ge b    \\
   0 &\quad\mbox{if}\quad x<a\\
   i &\quad\mbox{if}\quad \frac{x-a}{b-a}\in [\frac{i}{2^\kappa},\frac{i+1}{2^\kappa}),\,i\in\Z[0,2^{\kappa}).\\   
    \end{array}
\right. 
\end{eqnarray}
So for $a\le x < b$ 
$$
\beta_{\kappa,a,b}(x)=\left\lfloor 2^{\kappa}\,\frac{x-a}{b-a}\right\rfloor ,
$$
and hence, for $a\le x \le b$,
\begin{equation}\label{E1}
a+(b-a)\,2^{-\kappa}\beta_{\kappa,a,b}(x)\le x\le a+(b-a)\,2^{-\kappa}(\beta_{\kappa,a,b}(x)+1).
\end{equation}
First we state the basic result on quantum counting due to 
Brassard, H{\o}yer, Mosca, and Tapp (2000). 
\begin{lemma}
\label{lem:2}
There is a constant $c > 0$ such that for all $n, N \in \N$ there is a 
quantum algorithm $A$ from $\mathcal{B}_{\infty,0}^N$ to $\R$  
such that $n_q(A)\le n$ and for each $f\in \mathcal{B}_{\infty,0}^N$ 
$$
e(S_N,A,f)\le c\,\left(\sqrt{S_N f}\,n^{-1}+n^{-2}\right) .
$$ 
\end{lemma}
\noindent {\bf Remark.} 
Throughout this paper we often use the same symbol for possibly different constants.
These constants are either absolute or may depend only on $p$ -- 
the summability parameter of the $L_p$-spaces considered 
(in all lemmas and theorems this is precisely described anyway by the order of the
quantifiers).
\begin{proof}
We refer to Brassard, H{\o}yer, Mosca, and Tapp (2000) for details of the algorithm,
its analysis and the resulting estimates. For us, there remains one detail to be verified.
Their algorithm makes use of the controlled application of the Grover iterate 
and assumes that 
an implementation of this procedure is available.
This means, roughly, if  $Y$ stands for the Grover iterate, we must be able to
implement an operation which maps  an element $\lg i\rs\lg k\rs$ to 
$(Y^k\lg i\rs)\lg k\rs$ (that is, different basis elements may be subject to 
different powers of $Y$). 
Since $Y$ involves a query call, it is not immediately clear, how this could be achieved 
within the rules delveloped 
in section 2, that is, in our model of computation and its way 
to use queries. So we supply the needed argument here. It is a simulation procedure, 
similar to the ones above. 

The parameters of the algorithm will be the following. It has one measurement, and
the query $Q$ is determined by 
\begin{eqnarray*}
&\quad m'=\lceil \log N\rceil, \quad m''=1, \quad m^*=\lceil \log n\rceil,\\
&m=m'+2m^*+2, \quad Z=\Z[0,N),\\
 &\tau:Z\to Z[0,2^{m'}) \quad\mbox{and}\quad
\beta:\{0,1\}\to \{0,1\} \quad\mbox{the identities}
\end{eqnarray*}
(recall that $K=\{0,1\}$). Let
$$
H_m=H_{m'}\otimes H_1\otimes H_{m^*}\otimes H_{m^*}, 
$$
and let the basis state
$$
\lg i\rs\lg x\rs\lg j\rs \lg k\rs 
$$
correspond to this splitting. Let $\Phi_{n,m^*}$  be the $n$-term quantum
Fourier transform on $m^*$ qubits,
$$
\Phi_{n,m^*}\lg k\rs=\left\{\begin{array}{lll}
   \frac{1}{\sqrt{n}}\sum_{y=0}^{n-1}e^{2\pi\imath ky/n}\lg y\rs
   & \mbox{if} \quad k<n     \\
  \lg k\rs &\mbox{otherwise.}
    \end{array}\right.
$$
Define $\Phi\in\mathcal{U}(H_m)$ by
$$
\Phi\lg i\rs\lg x\rs\lg j\rs \lg k\rs=\lg i\rs\lg x\rs\lg j\rs (\Phi_{n,m^*}\lg k\rs).
$$
Furthermore, let $V_0\in\mathcal{U}(H_{m'})$ be the Walsh-Hadamard transform $W_N$,
if $N$ is a power of 2, and let $V_0=\Phi_{N,m'}$, if not. Define 
$X_0\in\mathcal{U}(H_{m'})$ by
$$
X_0\lg i\rs=\left\{\begin{array}{rll}
  -\lg i\rs &\mbox{if} \quad i=0 \\
  \lg i\rs & \mbox{otherwise,}    \\
    \end{array}
\right.
$$ 
and unitary transforms on $H_m$ by
\begin{eqnarray*}
V\lg i\rs\lg x\rs\lg j\rs \lg k\rs&=&(V_0\lg i\rs)\lg x\rs\lg j\rs \lg k\rs,\\
X\lg i\rs\lg x\rs\lg j\rs \lg k\rs&=&
\left\{\begin{array}{rll}
(X_0\lg i\rs)\lg x\rs\lg j\rs \lg k\rs & \mbox{if} \quad j<k   \\
 \lg i\rs\lg x\rs\lg j\rs \lg k\rs  & \mbox{otherwise,}    \\
    \end{array}
\right.\\ 
T\lg i\rs\lg x\rs\lg j\rs \lg k\rs&=&
\left\{\begin{array}{rll}
(-1)^{x+1} \lg i\rs\lg x\rs\lg j\rs \lg k\rs  &  \mbox{if} \quad j<k  \\
  \lg i\rs\lg x\rs\lg j\rs \lg k\rs &\mbox{otherwise,}  \\
    \end{array}
\right.\\
C\lg i\rs\lg x\rs\lg j\rs \lg k\rs&=&\lg i\rs\lg x\rs\lg j\oplus 1\rs \lg k\rs. 
\end{eqnarray*}
Now we define the algorithm as follows. For $f\in \mathcal{B}_{\infty,0}^N$ set
$$
Y_f=CVXV^{-1}Q_f T Q_f.
$$
The unitary transform of the algorithm is given by 
$$
\Phi^{-1}Y_f^{n-1}\Phi V.
$$
The initial state is 
$$
b=\lg 0\rs\lg 0\rs\lg 0\rs\lg 0\rs.
$$
Let us now follow the action of the algorithm. The element $b$ is transformed by $\Phi V$ 
into
$$
(V_0\lg 0\rs)\lg 0\rs\lg 0\rs (\Phi_{n,m^*}\lg 0\rs).
$$
Note that this vector is a linear combination of basis states of the form
$$
\lg i\rs\lg 0\rs\lg 0\rs\lg k\rs
$$
with $i<N$ and $k<n$.
Next consider the application of $Y_f$ to a basis state of the form
\begin{equation}
\label{N2}
\lg i\rs\lg 0\rs\lg j\rs\lg k\rs
\end{equation}
with $i<N$ and $k<n$. First we assume $j<k$. Then $Q_f T Q_f$ produces
$$
(-1)^{f(i)+1}\lg i\rs\lg 0\rs\lg j\rs\lg k\rs.
$$
After the application of $CVXV^{-1}$ we get 
$$
(-1)^{f(i)+1}(V_0 X_0 V_0^{-1}\lg i\rs)\lg 0\rs\lg j\oplus 1\rs\lg k\rs,
$$
which is a linear combination of vectors of the form
$$
\lg i'\rs\lg 0\rs\lg j+1\rs\lg k\rs
$$ 
with $i'<N$.
If $j\ge k$, the application of $Y_f$ to (\ref{N2}) gives
$$
\lg i\rs\lg 0\rs\lg j\oplus 1\rs\lg k\rs.
$$
 It is now clear that $Y_f=CVXV^{-1}Q_f T Q_f$ 
realizes the Grover iterate on the first component if $j<k$, that $Y_f^{n-1}$ is the 
controlled (by $k$) application of it and the whole algorithm, considered just on the 
first and last component $\lg i\rs\lg k\rs$, is the algorithm "Est\_Amp" of 
Brassard, H{\o}yer, Mosca, and Tapp (2000), if we define $\phi$ on the measured state
$$
\lg y\rs=\lg i\rs\lg x\rs\lg j\rs\lg k\rs
$$
as 
$$\phi(y)=\sin^2\left( \pi \frac{k}{n}\right).
$$
The required estimate (with a concrete value of the constant) is contained in Theorem 12 of
that paper. Since our implementation requires $2n$ queries, we rescale $n$ and
modify the constant appropriately.
\end{proof}
The next result is essentially a translation of Lemma \ref{lem:2} into
the setting of $\mathcal{B}_{\infty,+}^N$. The idea of using comparison queries is due 
to Abrams and Williams (1999). 
\begin{lemma}
\label{lem:3}
There is a constant $c > 0$ such that for all $\nu,n, N \in \N$ there is a 
quantum algorithm $A$ from $\mathcal{B}_{\infty,+}^N$ to $\R$ 
such that $n_q(A)\le \nu n$ and for each $f\in \mathcal{B}_{\infty,+}^N$
$$
e(S_N,A,f,2^{-\nu})\le c\,\left(\sqrt{S_N f}\,n^{-1}+n^{-2}\right) .
$$ 
\end{lemma}
\begin{proof}
Let $\kappa\in\N$ be such that  $2^\kappa \ge n^2$ and put $N_0=N\,2^\kappa$. 
We shall apply Corollary \ref{cor:1} with $F=\mathcal{B}_{\infty,+}^N$
and $\wt{F}=\mathcal{B}_{\infty,0}^{N_0}$.
Let
$\wt{A}$ be any algorithm from $\mathcal{B}_{\infty,0}^{N_0}$ to $\R$ with one measurement, 
 which satisfies the conclusion of Lemma \ref{lem:2} with $n_q(\wt{A}):=\wt{n}\le n$. Let
$\wt{A}$ be given by
$$
\wt{A}=(\wt{A}_0,\wt{b},\wt{\varphi}),\quad \wt{A}_0=(\wt{Q},(\wt{U}_j)_{j=0}^{\wt{n}}),
$$  
with
$$
\wt{Q}=(\wt{m},\wt{m}',\wt{m}'',\wt{Z},\wt{\tau},\wt{\beta}),
$$
where $\wt{Z}\subset\Z[0,2^{\wt{m}'})$, $\wt{\tau}:\wt{Z}\to\Z[0,N_0)$, and
$\wt{\beta}:\{0,1\}\to\Z[0,2^{\wt{m}''})$. 
We identify 
$$
\Z[0,N_0)= \Z[0,N)\times\Z[0,2^\kappa)
$$
and write correspondingly for $z\in \wt{Z}$,
\begin{equation}\label{20A}
\wt{\tau}(z)=(i(z),y(z)).
\end{equation}
Now let
 $\beta=\beta_{\kappa,0,1}$ as defined in (\ref{N1}). For each $f\in \mathcal{B}_{\infty,+}^N$
define $\Gamma(f)\in \mathcal{B}_{\infty,0}^{N_0}$ by setting for 
$(i,y)\in\Z[0,N)\times\Z[0,2^\kappa)=\Z[0,N_0)$
\[
\Gamma(f) (i,y) =
\left\{
    \begin{array}{llll}
1 & {\rm if}\quad y<\beta (f(i))\\
0 & {\rm otherwise}.
    \end{array}\right.
\]
Note that
\begin{eqnarray*}
\vert \{y : \Gamma(f) (i,y) = 1\} \vert =  \beta (f(i))  ,
\end{eqnarray*}
and consequently
$$
S_{N_0}\Gamma(f)=N^{-1}2^{-\kappa}\sum_{i=0}^{N-1}\beta (f(i)).
$$
By (\ref{E1}),
$$
S_{N_0}\Gamma(f)\le S_Nf\le S_{N_0}\Gamma(f)+2^{-\kappa}\le S_{N_0}\Gamma(f)+n^{-2}.
$$
The mapping $\Gamma:f\to\Gamma(f)$ is easily seen to be of the form (\ref{F1}) (with
$\eta(i,y)=i$ and $\beta$ as defined above). By Corollary \ref{cor:1}
there is an algorithm $A$ on $\mathcal{B}_{\infty,+}^N$ such that 
$n_q(A)= 2 n_q(\wt{A})$ and $A(f)=\wt{A}(\Gamma(f))$. To estimate the error of $A$,
fix any $f\in \mathcal{B}_{\infty,+}^N$ and 
let $\zeta$ be a random variable with distribution
$\wt{A}(\Gamma(f))$. Then, with probability at least 3/4,
\begin{eqnarray*}   
|S_Nf-\zeta|
&\le&|S_Nf-S_{N_0}\Gamma(f)|+|S_{N_0}\Gamma(f)-\zeta|\\
&\le&n^{-2}+c\,\left(\sqrt{S_{N_0}\Gamma(f)}\,n^{-1}+n^{-2}\right)\\
&\le&c'\,\left(\sqrt{S_N f}\,n^{-1}+n^{-2}\right).\\
\end{eqnarray*}
Now we use Lemma \ref{lem:2e} to boost the success probability by repeating
$A$ $c_1 \nu$ times, where $c_1=\lceil 8/\log\,e\rceil$, 
and computing the median, which 
gives the desired error estimate
$$
e(S,A^*,f,2^{-\nu})\le c'\,\left(\sqrt{S_N f}\,n^{-1}+n^{-2}\right)
$$
for the algorithm $A^*=\psi_0(A^{c_1\nu})$,
whose number of queries is bounded by $2c_1 \nu n$. A scaling of
$n$ at the expense of enlarging the constant gives the result as required.
\end{proof}
Now we are ready to estimate the numbers $e_n^q(S_N,\mathcal{B}_p^N)$. Note
that this is nontrivial only when $n<N$. For $n\ge N$ a classical computer suffices, 
or, to put it more formally into our framework, we have 
$e_n^q(S_N,\mathcal{B}_p^N)=0$, since with $N$ queries (and a suitable 
number of qubits) the sum can be determined up to each degree of precision by
e.g.\ simulating a classical computation. 

The following
is the main result of this section. For the sake of later reference
we also include the already known case $p=\infty$ due to 
Brassard, H{\o}yer, Mosca, and Tapp (2000),  which we deduce formally from the case $2<p<\infty$,
but which is, in fact, an immediate consequence
of the previous two lemmas.
\begin{theorem}
\label{theo:1}
Let $1 < p \le \infty$. 
Then there is a constant $c > 0$ such that for all $n, N \in \N$, $n>2$   
$$
e_n^q(S_N,\mathcal{B}_p^N)\le c\,
   \left \{\begin{array}{lll}
       n^{-1} & {\rm for} & p > 2 \\
       n^{-1} \log^{3/2}n  \log \log n &{\rm for} & p = 2\\                			  
        n^{-2 (1-1/p)}&{\rm for} & p < 2.\\
   \end{array}\right.
$$ 
\end{theorem}
\begin{proof} Let $1<p<\infty$. Fix $k\in \N_0$ (to be specified later) and define for $f\in L_p^N$
$$
{\cal I}_f^k=\left\{i\in\Z[0,N)\,\big|\,|f(i)|\ge 2^k\right\},
$$
for $\sigma=0,1$,
$$
{\cal J}_f^{0,\sigma}=\left\{i\in\Z[0,N)\,\big|\,0\le (-1)^\sigma f(i)<1\right\},
$$
and for  $\ell=1,\dots,k$,
$$
{\cal J}_f^{\ell,\sigma}=\left\{i\in\Z[0,N)\,\big|\,2^{\ell-1}
\le (-1)^\sigma f(i)<2^\ell\right\}.
$$
Note that 
$$
N^{-1}2^{pk}|{\cal I}_f^k|\le 
\frac{1}{N}\sum_{i\in {\cal I}_f^k}|f(i)|^p\le \|f\|_{L_p^N}^p,
$$
hence
\begin{equation}
\label{A6}
|{\cal I}_f^k|\le N\,2^{-pk}\|f\|_{L_p^N}^p.
\end{equation}
H\"older's inequality together with (\ref{A6}) gives
\begin{eqnarray}
\Big |\frac{1}{N}\sum_{i\in {\cal I}_f^k}f(i)\Big |
&\le&
\Big(\frac{1}{N}\big| {\cal I}_f^k\big|\Big)^{1/p'}
\Big(\frac{1}{N}\sum_{i\in {\cal I}_f^k}|f(i)|^p\Big)^{1/p}
\nonumber\\
&\le& 
2^{-pk/p'}\|f\|_{L_p^N}^{p/p'}\|f\|_{L_p^N}=2^{-(p-1)k}\|f\|_{L_p^N}^p,\label{A12a}
\end{eqnarray}
where $1/p+1/p'=1$. Furthermore,
\begin{equation}
\label{A4}
\frac{1}{N}\sum_{1\le\ell\le k \atop \sigma=0,1}2^{p(\ell-1)}|{\cal J}_f^{\ell,\sigma}|
\le\|f\|_{L_p^N}^p,
\end{equation}
which gives, in particular,
\begin{equation}
\label{A5}
|{\cal J}_f^{\ell,\sigma}|\le N\, 2^{-p(\ell-1)}\|f\|_{L_p^N}^p \quad (\ell\ge 1).
\end{equation}
 Now define $g_{f}^{\ell,\sigma}\in \mathcal{B}_{\infty,+}^N$
for $0\le\ell\le k$, $\sigma=0,1$,
\[
g_{f}^{\ell,\sigma} (i) =
                     \left\{
                      \begin{array}{lll}
                       (-1)^\sigma 2^{-\ell} f(i) & i \in {\cal J}_f^{\ell,\sigma}\\
                       0                       & {\rm otherwise.}
		      \end{array}
                     \right.
\]
Consequently, $0\le g_{f}^{\ell,\sigma} \le 1$, so 
$g_{f}^{\ell,\sigma}\in \mathcal{B}_{\infty,+}^N$. Clearly,
\begin{equation}
\label{A10a}
S_N g_{f}^{\ell,\sigma}\le N^{-1}|{\cal J}_f^{\ell,\sigma}|
\end{equation}
and
\begin{eqnarray}
S_Nf &=& \frac{1}{N}\Big(\sum_{0\le\ell\le k \atop \sigma=0,1}
\sum_{i\in {\cal J}_f^{\ell,\sigma}}f(i)+\sum_{i\in {\cal I}_f^k}f(i)\Big)\nonumber\\
&=&\sum_{0\le\ell \le k \atop \sigma=0,1}(-1)^\sigma 2^\ell S_N g_{f}^{\ell,\sigma}
+\frac{1}{N}\sum_{i\in {\cal I}_f^k}f(i).\label{A7}
\end{eqnarray}
Now the idea is to compute $S_N g_{f}^{\ell,\sigma}$ by the algorithm from Lemma 
\ref{lem:3} for all $\ell$ and $\sigma$, and from the results (in a classical
way) the first sum of equation (\ref{A7}).  Fix $\nu_\ell ,n_\ell\in\N$
(to be specified later) and let, according to Lemma \ref{lem:3}, $\wt{A}_\ell$ be an algorithm 
on $\mathcal{B}_{\infty,+}^N$ such that 
$n_q(\wt{A}_\ell)\le\nu_\ell n_\ell$ and for all $g\in \mathcal{B}_{\infty,+}^N$,
\begin{equation}
\label{A8}
e(S_N,\wt{A}_\ell,g,2^{-\nu_\ell})\le c\,\left(\sqrt{S_N g}\,n_\ell^{-1}+n_\ell^{-2}\right).
\end{equation}
 We define for $x\in\R$, $\sigma=0,1$,
$$
\rho_{0,\sigma}(x)=
\left\{\begin{array}{ll}
 (-1)^\sigma x \quad & {\rm if}\quad 0\le (-1)^\sigma x<1     \\
  0 &  {\rm otherwise,}   \\
    \end{array}
\right.
$$
and for $\ell=1, \dots,k-1$,
$$
\rho_{\ell,\sigma}(x)=
\left\{\begin{array}{ll}
 (-1)^\sigma 2^{-\ell}x \quad & {\rm if}\quad 2^{\ell-1}\le (-1)^\sigma x<2^\ell      \\
  0 &  {\rm otherwise.}   \\
    \end{array}
\right.
$$
Furthermore, we let $\eta$ be the identity on $\Z[0,N)$.
Then for each $f\in L_p^N$,
$$ 
g_{f}^{\ell,\sigma}=\rho_{\ell,\sigma}\circ f \circ \eta.
$$
By Corollary \ref{cor:1} there is an algorithm
$A_{\ell,\sigma}$ on $L_p^N$ with 
$$
n_q(A_{\ell,\sigma})=n_q(\wt{A}_\ell)
$$ 
and 
\begin{equation}
\label{A10}
A_{\ell,\sigma}(f)= \wt{A}_\ell (g_{f}^{\ell,\sigma})
\end{equation}
for all $f\in L_p^N$. We define $A$ as being composed of $A_{\ell,\sigma}$
 (in the sense of (\ref{A0})) as follows: 
$$
A=\sum_{0\le\ell \le k \atop \sigma=0,1}(-1)^\sigma 2^\ell A_{\ell,\sigma}.
$$
To estimate the error of $A$, fix any $f\in L_p^N$
and let 
$\{\zeta_{\ell,\sigma}\,|\, 0\le\ell\le k,\,\sigma=0,1\}$
 be independent random 
variables with distribution $A_{\ell,\sigma}(f)$ respectively.
Define
\begin{equation}\label{A9}
\zeta=\sum_{0\le\ell \le k \atop \sigma=0,1}
                          (-1)^\sigma 2^\ell \zeta_{\ell,\sigma}. 
\end{equation}
It follows from Lemma \ref{lem:2a} that
\begin{equation}\label{A9a}
A(f)=\mbox{dist}(\zeta).
\end{equation}
 By (\ref{A8}) and (\ref{A10}),
we have, with probability at least $1-2^{-\nu_\ell}$,
$$
|S_N g_{f}^{\ell,\sigma}-\zeta_{\ell,\sigma}|\le
c\,\left(\sqrt{S_N g_{f}^{\ell,\sigma}}\,n_\ell^{-1}+n_\ell^{-2}\right),
$$
and therefore, with probability at least 
$1-2\sum_{\ell=0}^k 2^{-\nu_\ell}$
$$
\big |\sum_{0\le\ell \le k \atop \sigma=0,1}(-1)^\sigma 2^\ell 
(S_N g_{f}^{\ell,\sigma}-\zeta_{\ell,\sigma})\big |
\le c\sum_{0\le\ell \le k \atop \sigma=0,1}
2^\ell \left(\sqrt{S_N g_{f}^{\ell,\sigma}}\,n_\ell^{-1}+n_\ell^{-2}\right),
$$
hence, by (\ref{A7}) and  (\ref{A9}),
$$
|S_Nf-\zeta|\le c\sum_{0\le\ell \le k \atop \sigma=0,1}
2^\ell \left(\sqrt{S_N g_{f}^{\ell,\sigma}}\,n_\ell^{-1}+n_\ell^{-2}\right)
+\Big |\frac{1}{N}\sum_{i\in {\cal I}_f^k}f(i)\Big |,
$$
which gives together with (\ref{A9a}), (\ref{A12a}), (\ref{A10a}) and  (\ref{A5}) 
\begin{eqnarray}
\lefteqn{
e(S_N,A,f,2\sum_{\ell=0}^k 2^{-\nu_\ell})
}
\nonumber\\
&\le&
c\sum_{0\le\ell \le k \atop \sigma=0,1}
2^\ell \left(\sqrt{S_N g_{f}^{\ell,\sigma}}\,n_\ell^{-1}+n_\ell^{-2}\right)
+2^{-(p-1)k}\|f\|_{L_p^N}^p \nonumber\\
&\le&
c\sum_{0\le\ell \le k \atop \sigma=0,1} 2^\ell\left(\sqrt{N^{-1}|{\cal J}_f^{\ell,\sigma}|}
\,n_\ell^{-1}+n_\ell^{-2}\right)
+2^{-(p-1)k}\|f\|_{L_p^N}^p 
\label{A11}\\
&\le&
2c\sum_{\ell =1}^k \left(2^{(1-p/2)\ell}n_\ell^{-1}\|f\|_{L_p^N}^{p/2}
+2^\ell n_\ell^{-2}\right) \nonumber\\
&&+\,2cn_0^{-1}+2^{-(p-1)k}\|f\|_{L_p^N}^p
\label{A12}
\end{eqnarray}
(recall the remark about constants after Lemma \ref{lem:2}). Moreover, we have
\begin{equation}
\label{A10b}
n_q(A)\le2\sum_{\ell=0}^{k}\nu_\ell n_\ell .
\end{equation}
Now we choose the parameters $k$, $\nu_\ell$ and $n_\ell$ in a suitable way
and prove the error estimates. First we consider the case
$2<p<\infty$. Here we put 
\begin{equation}
\label{A12b}
k=\left\lceil\frac{1}{p-1}\log n\right\rceil.
\end{equation}
Define, furthermore, $\nu_\ell=\lceil2\log(\ell+1)\rceil+4$, hence
\begin{equation}
\label{A13}
2\sum_{\ell=0}^k 2^{-\nu_\ell} 
\le \frac{1}{8}\sum_{\ell=0}^k (\ell+1)^{-2}<
\frac{1}{4}.
\end{equation}
Finally, let
\begin{equation}
\label{A14}
n_\ell=\left\lceil 2^{(1/2-p/4)\ell}n\right\rceil.
\end{equation}
This together with (\ref{A10b}) implies
\begin{equation}
\label{A14a}
n_q(A)\le 2\sum_{\ell=0}^k (\lceil2\log(\ell+1)\rceil+4)\left\lceil 
2^{(1/2-p/4)\ell}n\right\rceil\le c_1 n
\end{equation}
for some constant $c_1>0$.
It follows from (\ref{A13}), (\ref{A12}), (\ref{A14})
and (\ref{A12b}) that
\begin{eqnarray}
\lefteqn{
e(S_N,A,f)
}\nonumber\\
&\le& 
e(S_N,A,f,2\sum_{\ell=0}^k 2^{-\nu_\ell})\nonumber\\
&\le&
c\sum_{\ell =1}^k \left(2^{(1/2-p/4)\ell}n^{-1}\|f\|_{L_p^N}^{p/2} 
+2^{p\ell/2} n^{-2}\right)+cn^{-1}+n^{-1}\|f\|_{L_p^N}^p\nonumber\\
&\le& c \left( n^{-1}\|f\|_{L_p^N}^{p/2} + 2^{(1-p/2)k}n^{-1}+n^{-1}+
n^{-1}\|f\|_{L_p^N}^p\right)\nonumber\\
&\le& cn^{-1}\max(\|f\|_{L_p^N}^p,1).\nonumber
\end{eqnarray}
Consequently,
$$
e(S_N,A,\mathcal{B}_p^N)\le cn^{-1},
$$
which together with  
 (\ref{A14a}) implies the desired result in the case
$2<p<\infty$. Note that the case $p=\infty$ also follows since 
$ \mathcal{B}_\infty^N \subseteq\mathcal{B}_p^N$ for any $p<\infty$.

Now we suppose $1<p<2$. Here we choose
\begin{equation}
\label{A16}
k=\left\lceil\frac{2}{p}\log n\right\rceil,
\end{equation}
\begin{equation}
\label{A17}
n_\ell=\left\lceil 2^{-(1/2-p/4)(k-\ell)}n\right\rceil,
\end{equation}
and $\nu_\ell=\lceil2\log(k-\ell+1)\rceil+4$, which implies that (\ref{A13}) 
holds again. Furthermore, by (\ref{A10b}), 
\begin{equation}
\label{A17a}
n_q(A)\le 2\sum_{\ell=0}^k (\lceil2\log(k-\ell+1)\rceil+4)
\left\lceil 2^{-(1/2-p/4)(k-\ell)}n\right\rceil\le c_1 n.
\end{equation}
We get from (\ref{A13}),
 (\ref{A12}),  (\ref{A17})
and (\ref{A16}) 
\begin{eqnarray}
\lefteqn{
e(S_N,A,f)
}\nonumber\\
&\le & 
c\sum_{\ell =1}^k \left(2^{(1-p/2)\ell+(1/2-p/4)(k-\ell)}n^{-1}\|f\|_{L_p^N}^{p/2} 
+2^{\ell+(1-p/2)(k-\ell)} n^{-2}\right)\nonumber\\
&&+\,cn_0^{-1}+2^{-(p-1)k}\|f\|_{L_p^N}^p\nonumber\\
&\le &
c\sum_{\ell =1}^k \left(2^{(1/2-p/4)(k+\ell)}n^{-1}\|f\|_{L_p^N}^{p/2} 
+2^{k-p(k-\ell)/2} n^{-2}\right)\nonumber\\
&&+\,c\,2^{(1/2-p/4)k}n^{-1}+2^{-(p-1)k}\|f\|_{L_p^N}^p\nonumber\\
&\le &
c\left(2^{(1-p/2)k}n^{-1}\|f\|_{L_p^N}^{p/2}+2^k n^{-2}
+2^{(1/2-p/4)k}n^{-1}+2^{-(p-1)k}\|f\|_{L_p^N}^p\right)\nonumber\\
&\le & c n^{-2(1-1/p)}\max(\|f\|_{L_p^N}^p,1).\label{A19}
\end{eqnarray}
Now (\ref{A17a}) and (\ref{A19}) yield the needed result.

Finally, we consider the case $p=2$. Here we define
\begin{equation}
\label{A20}
n_\ell\equiv n_0=\left\lceil n\,(\log n)^{-1}(\log\log n)^{-1}\right\rceil
\end{equation}
(recall that we assumed $n>2$, so $n_0$ is well-defined and $n_0\ge 1$), 
furthermore
\begin{equation}
\label{A21}
k=\left\lceil \log n_0\right\rceil
\end{equation}
and
\begin{equation}
\label{A22}
\nu_\ell\equiv\nu_0=\left\lceil \log (k+1)\right\rceil+3.
\end{equation}
It follows that
\begin{equation}
\label{A22a}
2\sum_{\ell=0}^k 2^{-\nu_\ell} \le \frac{1}{4}\sum_{\ell=0}^k\frac{1}{k+1}\le\frac{1}{4}
\end{equation}
and, by (\ref{A10b}),
\begin{equation}
\label{A23}
n_q(A)\le 2(k+1)\nu_0 n_0\le c_1 n.
\end{equation}
By (\ref{A22a}) and (\ref{A11}), the error satisfies 
$$
e(S_N,A,f)
\le
c\sum_{1\le\ell \le k \atop \sigma=0,1} 2^\ell\sqrt{N^{-1}|{\cal J}_f^{\ell,\sigma}|}
\,n_\ell^{-1}+c\sum_{\ell=1}^{k}2^\ell n_\ell^{-2}
+cn_0^{-1}(\|f\|_{L_2^N}^2+1).
$$
H\"older's inequality, applied to the first sum, gives 
\begin{eqnarray*}
e(S_N,A,f)&\le&
c\,(2k)^{1/2}\Big (N^{-1}\sum_{1\le\ell \le k \atop \sigma=0,1} 2^{2\ell}
|{\cal J}_f^{\ell,\sigma}|\Big )^{1/2} n_0^{-1}\\
&&+\, c\sum_{\ell=1}^k 2^\ell n_0^{-2}
+cn_0^{-1}(\|f\|_{L_2^N}^2+1),
\end{eqnarray*}
and by (\ref{A4}), (\ref{A20}), and (\ref{A21}), we finally get
\begin{eqnarray*}
e(S_N,A,f)&\le& c\,\big(k^{1/2}n_0^{-1}\|f\|_{L_2^N}+2^k n_0^{-2}
+n_0^{-1}(\|f\|_{L_2^N}^2+1)\big)\\
&\le& cn^{-1}\log^{3/2}n\,\log\log n\,\max(\|f\|_{L_2^N}^2,1).
\end{eqnarray*}
This implies the statement for $p=2$.
\end{proof}
{\bf Remark.} Since quantum algorithms are not linear, the statement of 
Theorem \ref{theo:1} does not give any information on $f\in L_p^N$ of norm greater
than one. Our proof, however, does. It shows that the algorithm developed for 
fixed $1<p<\infty$ and $n,N\in\N$ has the property that for all $f\in L_p^N$

\[
e(S_N,A,f)\le c
\left\{\begin{array}{lll}
   n^{-1}\max(\|f\|_{L_p^N}^p,1) &\mbox{if} \quad 2<p<\infty  \\
    n^{-1}\log^{3/2}n\,\log\log n\,\max(\|f\|_{L_2^N}^2,1)&\mbox{if} \quad p=2    \\
   n^{-2(1-1/p)}\max(\|f\|_{L_p^N}^p,1) & \mbox{if} \quad 1<p<2.   \\  
    \end{array}
\right. 
\]

\section{Lower Bounds}

In this section  we derive lower bounds on the quantities $e_n^q(S,F)$ 
first in the general setting and then for  $F=\mathcal{B}_p^N$, $S=S_N$. 
Let $D$ and $K$ be nonempty sets, let
$L\in \N$ and let to each $u=(u_0,\dots,u_{L-1})\in\{0,1\}^L$ an 
$f_u\in \mathcal{F}(D,K)$ be assigned such that the following 
is satisfied:
\\ \\
{\bf Condition (I):} For each $t\in D$ there is an $\ell$, $0\le \ell\le L-1$, such that $f_u(t)$ 
depends only on $u_\ell$, in other words, for $u,u'\in\{0,1\}^L$, 
$u_\ell=u'_\ell$ implies $f_u(t)=f_{u'}(t)$.
\\ \\ 
This type of function system will play a key r\^ole in our lower bound proofs.
Condition (I) is easily seen to be equivalent to the following 
\\ \\
{\bf Condition (Ia):} There are functions $g_0,g_1\in \mathcal{F}(D,K)$ and a decomposition
$D=\bigcup_{\ell=0}^{L-1}D_\ell$ with $D_\ell\cap D_{\ell'}=\emptyset$ ($l\ne l'$)
such that for $t\in D_\ell$
\[
f_u(t)=
\left\{\begin{array}{lll}
   g_0(t)&{\rm if}& u_\ell=0  \\
    g_1(t)&{\rm if}& u_\ell=1. \\
    \end{array}
\right. 
\] \\
The first result is based on the polynomial method by Beals, Buhrman, Cleve, and Mosca (1998)
and extends their Lemma 4.1 to our general setting.
\begin{lemma}
\label{lem:4}
 Let $L\in\N$, let $(f_u)_{u\in\{0,1\}^L}\subseteq\mathcal{F}(D,K)$
be a system of functions satisfying condition (I), and 
let $A$ be a quantum algorithm on $\mathcal{F}(D,K)$ without measurement, $m=m_q(A)$,
$n=n_q(A)$. Then for all $x,b\in\Z[0,2^m),$
$A_{f_u}(x,b)$ (defined in (\ref{B1a}) and (\ref{14A2})), considered as a function of $u$, 
is a complex multilinear polynomial in 
the variables $u_0,\dots,u_{L-1}$ of degree at most $n$.
\end{lemma}
\begin{proof}
Let 
$$
A=(Q,(U_j)_{j=0}^n),\quad
Q=(m,m',m'',Z,\tau,\beta).
$$
Fix $b\in \Z[0,2^m)$ and define $w_j$ and $p_j(x,u)$ for $j=0,\dots,n$ by
$$
w_j=U_j Q_{f_u}U_{j-1}Q_{f_u}\dots U_1 Q_{f_u}U_0 b=  \sum_{x\in\Z[0,2^m)}p_j(x,u)\lg x \rs.
$$
Then 
\begin{equation}
\label{M4}
p_n(x,u)=A_{f_u}(x,b).
\end{equation}
Of course, $p_0(x,u)$ are constants, so polynomials of degree 0 in $u$. Now we
proceed by induction over $j$. Assume that for some $j$, $0\le j<n$, the 
$p_j(x,u)$ are polynomials of degree $\le j$ in $u$. Define $q_j(x,u)$ by 
$$
Q_{f_u}w_j=\sum_{x\in\Z[0,2^m)}q_j(x,u)\lg x \rs.
$$
Since
$$
Q_{f_u}w_j=Q_{f_u}\sum_{x\in\Z[0,2^m)}p_j(x,u)\lg x \rs
= \sum_{x\in\Z[0,2^m)}p_j(x,u)\,Q_{f_u}\lg x \rs,
$$
and since $Q_{f_u}$ is a bijection on the basis states, we get
$$
q_j(x,u)=p_j(Q_{f_u}^{-1}x,u).
$$
Now fix $x\in\Z[0,2^m)$. Represent $\lg x \rs$ as 
$\lg i \rs\lg y \rs\lg z \rs$ with $i\in \Z[0,2^{m'})$, $y\in \Z[0,2^{m''})$
and $z\in \Z[0,2^{m-m'-m''})$. According to the query definition (\ref{B1}),
we have $Q_{f_u}\lg x \rs=\lg i \rs\lg y \rs\lg z \rs$ if $i\not\in Z$. Hence, in this case 
$q_j(x,u)=p_j(x,u)$, so $\deg q_j(x,\,\cdot\,)\le j$. If $i\in Z$,
$$
Q_{f_u}^{-1}\lg x \rs= \lg i\rs\lg y\ominus\beta(f_u(\tau(i)))\rs\lg z\rs.
$$
Let, according to condition (I) above, $\ell$ be such that  $0\le\ell\le L-1$ and
$f_u(\tau(i))$ depends only on $u_\ell$. We denote $f_u(\tau(i))=s_0$ for $u_\ell=0$ 
and $f_u(\tau(i))=s_1$ for $u_\ell=1$. 
It follows that 
$$
Q_{f_u}^{-1}\lg x \rs =\lg i\rs\lg y\ominus\beta(f_u(\tau(i)))\rs\lg z\rs=
\left\{\begin{array}{lll}
 \lg i\rs\lg y\ominus\beta(s_0)\rs\lg z\rs:=x_0  & {\rm if}   &u_\ell=0 \\
 \lg i\rs\lg y\ominus\beta(s_1)\rs\lg z\rs:=x_1  & {\rm if}   &u_\ell=1.  \\
    \end{array}
\right. 
$$
Consequently, 
$$
q_j(x,u)=p_j(Q_{f_u}^{-1}x,u)=(1-u_\ell) p_j(x_0,u) + u_\ell p_j(x_1,u),
$$
which implies  $\deg q_j(x,\,\cdot\,)\le j+1$.  Now
$$
w_{j+1}=U_{j+1}Q_{f_u}w_j= U_{j+1}\sum_{y\in\Z[0,2^m)}q_j(y,u)\lg y \rs,
$$
which gives
$$
p_{j+1}(x,u)=\sum_{y\in\Z[0,2^m)}U_{j+1}(x,y)q_j(y,u),
$$
where $(U_{j+1}(x,y))_{x,y\in\Z[0,2^m)}$ is the matrix of the transformation 
$U_{j+1}$ in the canonical basis. Since the $U_{j+1}(x,y)$ are scalars 
not depending on $u$, and since $\deg q_j(x,\,\cdot\,)\le j+1$, it follows that 
$\deg p_{j+1}(x,\,\cdot\,)\le j+1$. This completes the
induction and shows that $\deg p_n(x,\,\cdot\,)\le n$. Now the lemma follows from
(\ref{M4}) and the observation that,
since the $u_i$ take only the values 0 and 1, we can replace any polynomial
by a multilinear one without changing its values on $\{0,1\}^L$. 
\end{proof}

\begin{corollary}
\label{cor:2}
Let $L\in\N$ and assume that  $(f_u)_{u\in\{0,1\}^L}\subseteq\mathcal{F}(D,K)$
satisfies condition (I). 
Let $A$ be a quantum algorithm from $\mathcal{F}(D,K)$ to a normed space $G$.
Then for each subset $C\subseteq G$,
$$
p(u)=A(f_u)\{C\}
$$
is a real multilinear polynomial of degree at most $2n_q(A)$.
\end{corollary}
\begin{proof}
This follows readily from Lemma \ref{lem:4} and relations (\ref{M1}) and (\ref{M3}).
\end{proof}

The next lemma is based on the results of Nayak and Wu (1999). To state it, 
we introduce some further notation. Define the function $\rho (L,\ell,\ell')$
for $L\in\N$, $0\le\ell\ne\ell'\le L$ by
\begin{equation}
\label{C1}
\rho (L,\ell,\ell')=\sqrt{\frac{L}{|\ell-\ell'|}}+
\frac{\min_{j=\ell,\ell'}\sqrt{j(L-j)}}{|\ell-\ell'|}.
\end{equation}
Note that $j(L-j)=(L/2)^2-(L/2-j)^2$, so this expression is minimized
iff $|L/2-j|$ is maximized.
For $u\in\{0,1\}^L$ set $|u|=\sum_{\ell=0}^{L-1}u_\ell$.
\begin{lemma}
\label{lem:5} There is a constant $c_0>0$ such that the following holds:
Let $D,K$ be nonempty sets, let 
$F\subseteq\mathcal{F}(D,K)$ be a set of functions, $G$ a normed space,
$S:F\to G$ a function, and $L\in\N$. Suppose 
$(f_u)_{u\in\{0,1\}^L}\subseteq\mathcal{F}(D,K)$ is a system of functions satisfying
condition (I). Let finally $0\le\ell\ne\ell'\le L$ and assume that 
\begin{equation}
\label{C2}
f_u\in F \quad {\rm whenever} \quad
|u|\in\{\ell,\ell'\}.
\end{equation}
 Then
\begin{equation}
\label{C3}
e_n^q(S,F)\ge \frac{1}{2}\min\big\{ \|S(f_u)-S(f_{u'})\|\,\big |\, |u|=\ell,\, 
|u'|=\ell'\big\}
\end{equation}
for all $n$ with
\begin{equation}
\label{C4}
n\le c_0\rho (L,\ell,\ell').
\end{equation}
\end{lemma}
\begin{proof} Nayak and Wu (1999, Theorem 1.1) showed that there is a constant $c>0$
such that for all $L\in\N$ and $0\le\ell\ne\ell'\le L$ the following holds:
If $p$ is an $L$-variate real polynomial such that 
$$
-1/4\le p(u)\le 5/4 \quad  \mbox{for all}\quad u\in\{0,1\}^L,
$$
$$
3/4\le p(u)\le 5/4 \quad {\rm if}\quad u\in\{0,1\}^L,\, |u|=\ell,
$$
and
$$
-1/4\le p(u)\le 1/4 \quad {\rm if}\quad u\in\{0,1\}^L,\, |u|=\ell',
$$
then 
\begin{equation}
\label{C11}
\deg p\ge c\rho (L,\ell,\ell'), 
\end{equation}
where $\rho$ was defined in (\ref{C1}).
Denote for $j=\ell,\ell'$
\begin{equation}
\label{C5}
G_j=\{S(f_u)\,|\,|u|=j\}
\end{equation}
and 
\begin{equation}
\label{C12}
\delta=d(G_\ell,G_{\ell'}),
\end{equation}
where for $X,Y\subseteq G$,
$$
d(X,Y)=\inf_{x\in X,\,y\in Y} \|x-y\|.
$$
(For $x\in G$ we write $d(x,G)$ instead of $d(\{x\},G)$.)
Now let $A$ be any quantum algorithm from $F$ to $G$ with
$n_q(A)=n$ and 
\begin{equation}
\label{C7}
e(S,A,F)<\delta/2.
\end{equation}
As we mentioned after the definition, a quantum algorithm on
$F$ is always also a quantum algorithm on $\mathcal{F}(D,K)$. 
For each $u\in\{0,1\}^L$, let 
$\zeta_u$ be a random variable with distribution $A(f_u)$. 
Define 
$$
p(u)=A(f_u)\{g\in G\,|\,d(g,G_\ell)<\delta/2\}=\Prm\{ d(\zeta_u,G_\ell)<\delta/2\}.
$$
It follows that 
\begin{equation}
\label{C8}
0\le p(u)\le 1
\end{equation}
and, by Corollary \ref{cor:2}, $p$ is a real polynomial satisfying
\begin{equation}
\label{C6}
\deg p \le 2n.
\end{equation}
Because of (\ref{C2}) and (\ref{C7}), we have for $|u|=\ell$,
\begin{eqnarray}
3/4 &\le&\Prm\{\|S(f_u)-\zeta_u\|<\delta/2\}\nonumber\\
&\le& \Prm\{d(\zeta_u,G_\ell)<\delta/2\}
=p(u).\label{C9}
\end{eqnarray}
On the other hand, for $|u|=\ell'$,
\begin{eqnarray}
1/4 &\ge&\Prm\{\|S(f_u)-\zeta_u\|\ge\delta/2\}\nonumber\\
&\ge& \Prm\{d(\zeta_u,G_{\ell'})\ge\delta/2\}\nonumber\\
&\ge& \Prm\{d(\zeta_u,G_{\ell})<\delta/2\}
=p(u).\label{C10}
\end{eqnarray}
From (\ref{C8} -- \ref{C10}) and (\ref{C11}), we infer
$$
2n\ge \deg p \ge c\rho(L,\ell,\ell').
$$
Now choose any $c_0<c/2$. Then $n\le c_0\rho (L,\ell,\ell')$ implies
$e_n^q(S,F)\ge \delta/2$, which, because of (\ref{C5}) and (\ref{C12}),
 is the same as (\ref{C3}).

\end{proof}
The following theorem is the main result of this section. The case $p=\infty$ is due to 
Nayak and Wu (1999), and the case $2\le p <\infty$ is a direct consequence.
For the sake of completeness we include this part in the proof below.
(Another reason for this is that we use a slightly more general
notion of query, so this way we formally check that their bound holds true
also for our model.) 
\begin{theorem}
\label{theo:2}
Let $1 \le p \le \infty$. 
Then there are constants $c_0,c_1,c_2 > 0$ such that for $n, N \in \N$,   
$$
e_n^q(S_N,\mathcal{B}_p^N)\ge c_2 
\left\{\begin{array}{lllll}
 n^{-2(1-1/p)} & {\rm if}& 1\le p<2 & {\rm and}& n\le c_0\sqrt{N}\\
    n^{-1}& {\rm if} &2\le p\le \infty  & {\rm and}&n\le c_1N.  \\
    \end{array}
\right. 
$$ 
\end{theorem}

\begin{proof} Let $c_0$ be the constant from Lemma \ref{lem:5}.
Let $1\le p<2$ and 
\begin{equation}
\label{D1}
n\le c_0\sqrt{N}.
\end{equation}
Define
\begin{equation*}
L=\left\lceil c_0^{-2}n^2\right\rceil,\quad \ell=0,\quad \ell'=1.
\end{equation*}
It follows from (\ref{D1}) that $1\le L\le N$. Moreover,
\begin{equation}
\label{D3}
n\le c_0\sqrt{L}=c_0\rho(L,\ell,\ell')
\end{equation}
and
\begin{equation}
\label{D4}
L <c_0^{-2}n^2 +1\le (c_0^{-2}+1)n^2.
\end{equation}
Put $M=\lfloor L^{-1}N\rfloor$. Hence $1\le M\le N$ and
\begin{equation}
\label{D5}
M\le L^{-1}N\le 2M.
\end{equation}
Define $\psi_j \quad (j = 0, \dots , L-1)$ 
by
\[
\psi_j (i) = 
  \left\{
   \begin{array}{lll}
(N/M)^{1/p} & {\rm if} \quad jM \le i < (j+1)M\\
0 & {\rm otherwise.}
   \end{array}
   \right.
\]
Note that $\psi_j\in \mathcal{B}_p^N$ and
$$
S_N\psi_j = (MN^{-1})^{1-1/p}.
$$
For each $u = (u_0, \dots , u_{L-1}) \in \{0,1\}^L$ define
\begin{equation}
\label{E6}
f_u = \sum_{j=0}^{L-1} u_j \psi_j.
\end{equation}
Since the functions $\psi_j$ have disjoint supports, the system
$(f_u)_{u\in\{0,1\}^L}$ satisfies condition (I). Lemma \ref{lem:5} 
and relation (\ref{D3}) together with (\ref{D5}) and (\ref{D4}) give
\begin{eqnarray*}
e_n^q(S_N,\mathcal{B}_p^N)&\ge& \frac{1}{2}\min\big\{ |S_N f_u-S_N f_{u'}|\,\big |\, 
|u|=0,\, |u'|=1\big\}\\
&=& \frac{1}{2} \left(MN^{-1}\right)^{1-1/p}\ge \frac{1}{2}(2L)^{-(1-1/p)}
\ge c_2 n^{-2(1-1/p)}
\end{eqnarray*}
for some constant $c_2>0$. This proves the statement in the first case.

Now we consider the case $2\le p\le\infty$. Since 
$\mathcal{B}_\infty^N\subset \mathcal{B}_p^N$
whenever $p<\infty$, it suffices to prove the lower bound for $p=\infty$.
We set $c_1=2^{-1}(c_0^{-1}+2)^{-1}$ and assume $n\le c_1 N$. Let
\begin{equation*}
L=2\left\lceil c_0^{-1}n+1\right\rceil, \quad \ell=L/2-1, \quad \ell'=\ell+1=L/2.
\end{equation*}
It follows that $L\ge 4$ and
\begin{equation}
\label{E3}
\rho(L,\ell,\ell')>\min_{j=l,l'}\sqrt{j(L-j)}=\sqrt{L^2/4-1}\ge c_0^{-1}n.
\end{equation}
Moreover, since $1\le n\le c_1 N$, we get  
\begin{equation}
\label{E4}
L=2\left\lceil c_0^{-1}n+1\right\rceil \le 2(c_0^{-1}n+2)\le 2(c_0^{-1}+2)n\le N.
\end{equation}
Now let $M=\left\lfloor L^{-1}N\right\rfloor$, then (\ref{D5}) holds again. Set
\[
\psi_j (i) = 
  \left\{
   \begin{array}{lll}
1& {\rm if} \quad jM \le i < (j+1)M\\
0 & {\rm otherwise}
   \end{array}
   \right.
\]
for $j = 0, \dots , L-1$, and let $f_u$ be again defined by (\ref{E6}). 
Clearly, $(f_u)_{u\in\{0,1\}^L}$ satisfies condition (I) and
 $f_u\in \mathcal{B}_\infty^N$ for all $u\in \{0,1\}^L$. 
Lemma \ref{lem:5} together with relations (\ref{E3}), (\ref{D5}) and (\ref{E4}) gives

\begin{eqnarray*}
e_n^q(S_N,\mathcal{B}_\infty^N)&\ge& \frac{1}{2}\min\big\{ |S_N f_u-S_N f_{u'}|\,\big |\, 
|u|=\ell,\, |u'|=\ell+1\big\}\\
&=& \frac{1}{2} MN^{-1}\ge \frac{1}{4L}
\ge c_2 n^{-1}
\end{eqnarray*}
for some $c_2>0$.
\end{proof}
{\bf Remark.} Comparing Theorem \ref{theo:2} with Theorem \ref{theo:1},
we see that matching upper and lower bounds were obtained except for the case 
of $1\le p<2$, $n\ge c_0\sqrt{N}$.
This case is settled in Heinrich and Novak (2001b).

\section{Integration in $L_p\big([0,1]^d\big)$}
Here we present an application of the summation results to integration of 
functions. Further results will be contained in Heinrich (2001). 
Let $1\le p \le\infty$, $d\in\N$, $D=[0,1]^d$ and let $L_p(D)$ denote the usual space 
of $p$-integrable with respect to the Lebesgue measure functions on $D$, equipped 
with the norm
$$
\|f\|_{L_p(D)}=\left(\int_D |f(t)|^p\,dt\right)^{1/p}
$$
if $p<\infty$ and
$$
\|f\|_{L_\infty (D)}=\mbox{ess\,sup}_{t\in D} |f(t)|.
$$
Let $I_d:L_p(D)\to \R$ be the integration operator, defined for $f\in L_p(D)$ by
$$
I_df=\int_D f(t)\,dt.
$$
In this chapter we will consider $G=\R$ and $S=I_d$. We want to integrate 
functions from the unit ball $\mathcal{B}(L_p(D))$ in the quantum model of computation 
developed in section 2. 
Strictly speaking, $L_p(D)$ consists of equivalence classes of functions being equal
almost everywhere. Hence, function values are not well-defined, in general. This changes, 
however, if we consider subsets of $L_p(D)$ which consist of continuous functions, or more
precisely, of equivalence classes which contain a (unique) continuous function. This
is how we shall approach the integration problem -- we study it for certain subsets
$\mathcal{E}\subset\mathcal{B}(L_p(D))$. We shall assume that $\mathcal{E}$ 
is an equicontinuous set of functions on $D$. Since $D$ is compact, equicontinuity is
equivalent to uniform equicontinuity, and the latter means that for each
$\e>0$ there is a $\delta>0$ such that for $s,t\in D$, $\|s-t\|_\infty\le\delta$ implies 
$|f(s)-f(t)|\le \e$ for all $f\in F$.
Note also that it follows readily from the Arzel\`{a}-Ascoli theorem that 
$\mathcal{E}\subset\mathcal{B}(L_p(D))$
is equicontinuous iff $\mathcal{E}$ is relatively 
compact in the space $C(D)$ of continuous functions on $D$, equipped with the sup-norm. 
(A similar approach was chosen in Novak, 1988, to discuss restricted Monte Carlo methods.)
\begin{theorem}
\label{theo:3}
Let $1\le p\le\infty$. Then there are constants $c_1,c_2>0$ such
 that for all $d,n\in \N$
\begin{eqnarray*}
c_1 n^{-1} &\le&  
\sup_{\mathcal{E}\subset\mathcal{B}(L_p(D))}
e_n^q(I_d,\mathcal{E}) \quad\le\quad  c_2 n^{-1}
\quad \quad\quad 2< p\le\infty \\
c_1 n^{-1} &\le& 
\sup_{\mathcal{E}\subset\mathcal{B}(L_2(D))}
e_n^q(I_d,\mathcal{E})\quad \le \quad c_2 n^{-1}\log^{3/2}n\log\log n\\
c_1 n^{-2(1-1/p)} &\le&  
\sup_{\mathcal{E}\subset\mathcal{B}(L_p(D))}
e_n^q(I_d,\mathcal{E}) \quad\le\quad  c_2 n^{-2(1-1/p)}\quad  1\le p<2,\\
\end{eqnarray*}
where the supremum is taken  over all equicontinuous subsets $\mathcal{E}$ of
$\mathcal{B}(L_p(D))$.
\end{theorem}
\begin{proof}
First we prove the upper bounds.
Let $\mathcal{E}\subset\mathcal{B}(L_p(D))$ be equicontinuous and let
$n\in\N$. For $k\in\N$ let 
\begin{eqnarray*}
D= \bigcup_{i=0}^{2^{dk}-1} D_i
\end{eqnarray*}
be the partition of $D$ into $2^{dk}$  congruent cubes of disjoint interior.
Let $s_i$ be the point in $D_i$ with the smallest Euclidean norm.
Let $P_k$ be the operator of piecewise constant interpolation with respect to the partition
$(D_i)_{i=0}^{2^{dk}-1}$ in the points  
$(s_i)_{i=0}^{2^{dk}-1}$ (to avoid ambiguity, if a point belongs to more than one of 
the sets $D_i$, we assign to it the value $f(s_i)$ for the smallest such $i$).
Due to the equicontinuity of $\mathcal{E}$ there is a $k\in\N$
such that
\begin{equation}
\label{K1}
\|f-P_kf\|_{L_\infty(D)}\le n^{-1}
\end{equation}
for all $f\in \mathcal{E}$. Fix this $k$ and put $N=2^{dk}$. It follows that
\begin{equation}
\label{G4}
\sup_{f\in\mathcal{E}} |I_df-I_d(P_kf)|\le n^{-1}.
\end{equation}
Moreover, defining $$
\Gamma:\mathcal{E}\to L_p^N \quad\mbox{by} 
\quad \Gamma(f)(i)=f(s_i)\quad (i=0,\dots,N-1),
$$
we get
\begin{equation}
\label{G5}
I_d(P_kf)=\frac{1}{N}\sum_{i=0}^{N-1}f(s_i)=S_N\circ\Gamma(f).
\end{equation}
Note that for $f\in\mathcal{E}\subset\mathcal{B}(L_p(D))$  
\begin{eqnarray*}
\left(\frac{1}{N}\sum_{i=0}^{N-1}|\Gamma(f)(i)|^p\right)^{1/p}&=&
\left(\int_D|P_kf(s)|^pds\right)^{1/p}=\|P_kf\|_{L_p(D)}\\
&\le& \|f\|_{L_p(D)}+\|f-P_kf\|_{L_p(D)}\\
&\le& \|f\|_{L_p(D)}+n^{-1}\le 2.
\end{eqnarray*}
Consequently, $\Gamma$ maps $\mathcal{E}$ into $2\mathcal{B}_p^N$. Lemma \ref{lem:2d},
Corollary \ref{cor:1} and 
 relations (\ref{G4}) and (\ref{G5}) imply
\begin{eqnarray*}
e_n^q(I_d,\mathcal{E})&\le& n^{-1}+e_n^q(S_N\circ\Gamma,\mathcal{E})
\le n^{-1}+e_n^q(S_N,2\mathcal{B}_p^N)\\
&=&n^{-1}+2e_n^q(S_N,\mathcal{B}_p^N),
\end{eqnarray*}
hence Theorem \ref{theo:1} yields the upper bounds.

To verify the lower bounds, fix a $\sigma$ with $0<\sigma<1$ and 
let $\psi$ be a continuous function on $\R^d$ with 
$$
{\rm supp}\, \psi\subseteq [0,1]^d, \quad 0\le\psi\le 1\quad \mbox{and}\quad I_d\psi=\sigma.
$$
Fix $n\in \N$ and choose $N=2^{dk}$ in such a way
that 
\begin{equation}
\label{H1}
c_0\sqrt{N}\ge n\quad\mbox{and}\quad c_1 N\ge n,
\end{equation}
 where $c_0$ and $c_1$ are the constants from Theorem \ref{theo:2}. 
Set
$$
\psi_i(t)=\psi(2^k(t-s_i))\quad (i=0,\dots,N-1),
$$
with the $s_i$ as in the preceding part of the proof. Consequently,
\begin{equation}
\label{G7}
I_d\psi_i=2^{-dk}I_d\psi=\sigma 2^{-dk}=\sigma N^{-1}.
\end{equation}
Define 
$$
\Gamma:\mathcal{B}_p^N\to L_p(D) \quad\mbox{by}\quad\Gamma(f)=\sum_{i=0}^{N-1}f(i)\psi_i.
$$
For $f\in \mathcal{B}_p^N$,
\begin{eqnarray*}
\|\Gamma(f)\|_{L_p(D)}&=&\int_D\sum_{i=0}^{N-1}|f(i)|^p|\psi_i(t)|^pdt
=\sum_{i=0}^{N-1}|f(i)|^p\int_D|\psi_i(t)|^pdt\\
&=&2^{-dk}\sum_{i=0}^{N-1}|f(i)|^p\int_D|\psi(t)|^pdt
\le N^{-1}\sum_{i=0}^{N-1}|f(i)|^p\le 1.
\end{eqnarray*}
We define $\mathcal{E}=\Gamma(\mathcal{B}_p^N)$, which is a subset of $\mathcal{B}(L_p(D))$.
Since the functions $\psi_i$ are continuous, and $|f(i)|\le N^{1/p}$ for all
$f\in \mathcal{B}_p^N$, the equicontinuity of $\mathcal{E}$ easily follows.
Furthermore,
$$
I_d\circ \Gamma(f)=I_d\sum_{i=0}^{N-1}f(i)\psi_i=\sum_{i=0}^{N-1}f(i)I_d\psi_i=
\sigma N^{-1}\sum_{i=0}^{N-1}f(i)=\sigma S_Nf.
$$
Lemma \ref{lem:2d} and Corollary \ref{cor:1} give
$$
\sigma e_n^q(S_N,\mathcal{B}_p^N)=e_n^q(\sigma S_N,\mathcal{B}_p^N)
\le e_n^q(I_d,\mathcal{E}),
$$
and the result follows from relation (\ref{H1}) and
 Theorem \ref{theo:2}.
\end{proof}
\section{Comments}
Our results were formulated in the language of 
information-based complexity theory -- the minimal error at given cost 
(number of function values, functionals etc., in our case queries). 
Lower bounds in terms of the number of queries mean the more  
that no algorithm can have better arithmetic (bit) cost. On the other hand, if we have
upper bounds on the number of queries, this does not necessarily mean a
corresponding estimate of the cost in the bit model. However, for the problems considered 
in this paper we encounter a situation which is largely parallel
to the experience in information-based complexity: As a rule, the developed 
algorithms, which are optimal in the query sense, show a similar behaviour (usually up 
to certain logarithmic terms) in their arithmetic (bit) cost. 
Let us have a closer look at our algorithms from this point of view. 

The bit cost of
one query of the type (\ref{J1}) we define to be $m'+m''$
(the number of bits to be processed). When we consider the bit cost, let us assume that 
both $N$ and $n$ are powers
of two, which is no loss of generality since the other cases can be reduced to that.
We also assume $n<N$, see the remarks before Theorem \ref{theo:1}.
 The algorithm from Lemma \ref{lem:2}  
makes one measurement and
can be implemented on $\mathcal{O}(\log N)$ qubits using $\mathcal{O}(n\,\log N)$ 
quantum gates.  The algorithm of Lemma 
\ref{lem:3} requires $\mathcal{O}(\log N)$ qubits, $\mathcal{O}(\nu n\log N)$
gates and makes $\mathcal{O}(\nu)$ measurements. Finally, the algorithm from 
Theorem \ref{theo:1}
needs $\mathcal{O}(\log N)$ qubits, $\mathcal{O}(n \log N)$ gates and 
$\mathcal{O}(\log n\,\log\log n)$ 
measurements for $p<\infty$
(one measurement if $p=\infty$). 

To discuss the algorithm of Theorem \ref{theo:3}, let us introduce the following quantity
for an equicontinuous
subset $\mathcal{E}\subset\mathcal{B}(L_p(D))$ and $\e>0$:
\begin{eqnarray*}
 \kappa(\mathcal{E},\e)&=& \min\,\{k\in \N\,\,|\\ 
&&|f(s)-f(t)|\le \e \,\, \mbox{whenever}\,\,
f\in\mathcal{E},\,s,t\in D, \|s-t\|_\infty\le 2^{-k}\}.
\end{eqnarray*}
Then for a given $\mathcal{E}\subset\mathcal{B}(L_p(D))$ we have to compute the mean of 
$N=2^{dk}$ numbers, where it suffices to take $k=\kappa(\mathcal{E},1/n)$.
If $N\le n$, this can be done with $\mathcal{O}(\log n)$ qubits,
$\mathcal{O}(N\log n)$ gates and one measurement (see the remarks before Theorem
\ref{theo:1}). If $n<N$,  
 this can be implemented on 
$\mathcal{O}(d\,\kappa(\mathcal{E},1/n))$ qubits, with
$\mathcal{O}(dn\,\kappa(\mathcal{E},1/n))$ gates and
$\mathcal{O}(\log n\,\log\log n)$ 
measurements for $p<\infty$ and one measurement for $p=\infty$.
(The constants in the $\mathcal{O}$-notation do not depend on $\mathcal{E}$ and $d$.)

Next let us compare the results obtained above to the classical deterministic
and Monte Carlo setting. We denote the respective quantities by $e_n^{det}$ and 
$e_n^{mc}$. This discussion is carried out in greater detail in Heinrich and Novak (2001a),
where also the related definitions and references can be found. The following table contains
the order of the respective quantities, that is, the behaviour up to constants. We also
omitted the additional logarithmic factor in the case $p=2$. Furthermore, we assume 
for the case
$\mathcal{B}_p^N$ that $n\le c_1N$, where in the classical settings, $c_1$ is any constant 
with
$0<c_1<1$, while in the quantum setting for $2\le p \le\infty$,
 $c_1$ is the constant from Theorem \ref{theo:2}. 
Moreover in the quantum setting for $1\le p<2$, we assume $n\le c_0\sqrt{N}$,
with $c_0$ from Theorem \ref{theo:2}, as well. Finally, when we write 
$\mathcal{B}_{L_p}$, we mean (in all three setttings) the supremum over all equicontinuous
subsets $\mathcal{E}\subset \mathcal{B}(L_p([0,1]^d))$ as in the previous section.

\[ 
\begin{array}{l|l|l|l}
&\quad  e_n^{det}  &\quad e_n^{mc} &\quad e_n^{q}\,\\ \hline
\mathcal{B}_p^N,\,2\le p\le \infty &\quad 1& \quad n^{-1/2}        & \quad n^{-1}\\
\mathcal{B}_p^N,\,1\le p<2 & \quad 1&\quad  n^{-1+1/p} & \quad  n^{-2+2/p}  \,\quad 
\\
\mathcal{B}_{L_p},\,2\le p\le \infty &\quad 1& \quad n^{-1/2 }  
& \quad n^{-1}\\
\mathcal{B}_{L_p},\,1\le p<2 &\quad 1& \quad n^{-1+1/p } & \quad n^{-2+2/p}
\end{array}
\]
The result on $\mathcal{B}_{L_p}$ in the randomized setting 
can be found in Heinrich (1993). The respective statement for the deterministic 
setting is easily derived using standard methods of information-based complexity theory.
A little further below we indicate the proof of a somewhat stronger result.

It might be illustrative to formulate the results in terms of complexity. Here
we impose the 
corresponding restrictions. We always assume $\e\le \e_0$ for some constant $\e_0>0$. 
In the quantum setting, the case $1< p<2$ holds only for
$N\ge c(1/\e)^{p/(p-1)}$, for some constant $c>0$. Again, the case $p=2$ holds up to
logarithmic terms.
\[ 
\begin{array}{l|l|l|l}
& \mbox{comp}_\varepsilon^{det}  & \mbox{comp}_\varepsilon^{mc} 
& \mbox{comp}_\varepsilon^q\\ \hline
\mathcal{B}_p^N,\,2\le p\le \infty &  N &  \min((1/\e)^2,N) &  \min((1/\e),N)\\
\mathcal{B}_p^N,\,1< p<2 &  N &  \min((1/\e)^{p/(p-1)},N)  
&  \min((1/\e)^{p/(2(p-1))},N) 
\\
\mathcal{B}_{L_p},\,2\le p\le \infty& \infty&  (1/\e)^2  
&  (1/\e) \\
\mathcal{B}_{L_p},\,1\le p<2 & \infty&  (1/\e)^{p/(p-1)}
 &  (1/\e)^{p/(2(p-1))}
\end{array}
\]
In the case $\mathcal{B}_{L_1}$ we have $\infty$ in all three settings. For $\mathcal{B}_1^N$
we have $N$ in both classical settings, while in the quantum setting our results give
the lower bound $\sqrt{N}$ and the (trivial) upper bound $N$. The question of the
correct order of complexity in this case is answered in Heinrich and Novak (2001b).

We see that for the problems considered here quantum algorithms reach  a 
quadratic speedup over classical randomized ones and -- at least as far as the
pure number of queries is concerned (disregarding the bit cost and number of qubits) --
an arbitrarily large speedup over classical deterministic algorithms.  Let us discuss this 
last point in some more detail and also address the bit issue again. Namely, we show
that there are equicontinuous sets $\mathcal{E}$ in $\mathcal{B}(L_\infty([0,1]))$ with arbitrarily 
slowly decreasing 
$e_n^{det}(I_1,\mathcal{E})$. More precisely, 
for any sequence $(\e_n)_{n\in\N}$ with 
\begin{equation}
\label{K3}
0<\e_n\le 1,\quad \e_{n+1}\le\e_n\le 2\e_{2n},\quad \mbox{and}\quad
\lim_{n\to\infty}\e_n=0
\end{equation}
there is an equicontinuous set 
$\mathcal{E}\subset \mathcal{B}(L_\infty([0,1]))$ such that for all $n\in \N$
\begin{equation}
\label{K2}
e_n^{det}(I_1,\mathcal{E})\ge\e_n/32.
\end{equation}
Indeed, we define $\mathcal{E}$ as the set of functions $f$ on $[0,1]$ such that for all 
$k\in\N$ and $s,t\in [0,1]$, $|s-t|\le2^{-k}$ implies $|f(s)-f(t)|\le\e_{2^k}$. 
Let 
\[
\psi(t)=
\left\{\begin{array}{lll}
  t & \mbox{if} \quad 0\le t\le 1/2   \\
  (1-t)& \mbox{if} \quad 1/2<t\le 1   \\
  0& \mbox{otherwise,}    \\
    \end{array}
\right. 
\]
and put for $k\in \N$ and $0\le i\le 2^k-1$ 
$$
\psi_{k,i}(t)=\e_{2^k}\psi(2^k(t-2^{-k}i)).
$$
It is easily checked that for any $\alpha_i\in\{-1,1\}\quad (i=0,\dots,2^k-1)$,
$$
\sum_{i=0}^{2^k-1}\alpha_i\psi_{k,i}\in \mathcal{E}
$$ 
and $I_1\psi_{k,i}=2^{-(k+2)}\e_{2^k}$. A standard argument from the deterministic 
setting of information-based complexity theory (see e.g.\ Novak, 1988, Prop.\ 1.3.5 b) 
yields 
$$
e_{2^{k-1}}^{det}(I_1,\mathcal{E})\ge \e_{2^k}/8\ge \e_{2^{k-1}}/16,
$$
which implies (\ref{K2}). Recall, on the other hand, that 
by Theorem \ref{theo:3},  \\ $e_n^q(I_1,\mathcal{E})\le cn^{-1}$.

Now let us turn to the bit cost. We show that an exponential speedup is possible.
Fix any $\gamma$ with $0<\gamma\le 1$. We choose $\e_1=1$ 
and $\e_n=(\log n)^{-\gamma}\,\,(n>1)$. This 
sequence satisfies 
(\ref{K3}). Let
$\mathcal{E}\subset\mathcal{B}(L_\infty([0,1]))$ be the corresponding set constructed above,
so that
$$
e_n^{det}(I_1,\mathcal{E})\ge (\log n)^{-\gamma}/32\quad (n>1),
$$
which means that for any $\e$ with $0<\e\le 1/32$ we need  at least 
$2^{(1/(32\e))^{1/\gamma}}$, that is,
 exponentially many operations 
to reach error $\e$ deterministically. By the construction of the set 
$\mathcal{E}$ we have
$$
\kappa(\mathcal{E},1/n)\le \lceil n^{1/\gamma}\rceil,
$$
which implies, by the discussion at the beginning of this section, that 
in the quantum setting, an error of $\e$ can be 
reached with $\mathcal{O}(1/\e)$ queries, one measurement, $\mathcal{O}((1/\e)^{1/\gamma})$
qubits and $\mathcal{O}((1/\e)^{1/\gamma+1})$ gates, that is, with polynomial total cost.

Finally we discuss a topic concerning the relations to information-based complexity.
A look at our notion of a query might lead to the impression that it covers only what
is called standard information, that is, function values of $f$, while in 
information-based complexity also more general types of information are considered
(e.g.\ arbitrary linear functionals or scalar products with certain
basis functions). This could be relevant not only in finite element methods, but also
in the case that function values are not well-defined. Let us show how our approach
covers also this situation.

So let $F$ and $K$ be  nonempty sets, $S:F\to G$ be a mapping from $F$
to a normed space $G$ and let
$\Lambda$ be a nonempty set of mappings from $F$ to $K$. We seek to approximate
$S$ again, but now the algorithm is supposed to use 
information about $f\in F$ of the form $\lambda(f)$ for $\lambda\in \Lambda$. 
Let us define a $\Lambda$-based quantum algorithm from $F$ to $G$ to be simply a 
quantum algorithm $A$ from $\mathcal{F}(\Lambda,K)$ to $G$. 
Introduce the mapping 
$$
\Psi:F\to \mathcal{F}(\Lambda,K)
$$
defined for $f\in F$ by
$$
 \Psi(f)(\lambda)=\lambda(f)\quad (\lambda\in \Lambda).
$$
The error of
$A$ at $f\in F$ is defined as follows. Let $\zeta$ be a random variable
with distribution $A(\Psi(f))$. Put 
\begin{equation*}
e(S,A,f,\theta)=\inf\left\{\varepsilon\,\,|\,\,\Prm\{\|S(f)-\zeta\|>\varepsilon\}\le\theta
\right\}.
\end{equation*}
Various further quantities like $e(S,A,F)$, $e_n^q(S,F,\Lambda)$ etc.\ can be defined on  
this basis as in section 2. The results of section 2 as well as the general results
of section 4 remain valid for this situation 
if formulated appropriately, that is, 
if applied to $A$ as an algorithm from $\mathcal{F}(\Lambda,K)$ to $G$. The resulting form of 
the unitary mappings associated with the query is worth mentioning: 
Let $Q$ be one of the queries 
being part of $A$. 
Since $A$ is an algorithm on $\mathcal{F}(\Lambda,K)$, its queries have the form
(\ref{J1}), where everything is as specified there except that 
$$\tau:Z\to \Lambda.$$ 
Let us denote $\lambda_i=\tau(i)$ for $i\in Z$. Then
an element $f\in F$ gives rise to the following unitary operator implementing the query 
\begin{equation*}
Q_{\Psi(f)}\lg i\rs\lg x\rs\lg y\rs=
\left\{\begin{array}{ll}
\lg i\rs\lg x\oplus\beta(\lambda_i(f))\rs\lg y\rs &\quad \mbox {if} \quad i\in Z\\
\lg i\rs\lg x\rs\lg y\rs & \quad\mbox{otherwise.} 
 \end{array}
\right. 
\end{equation*}

\medskip 
\noindent
{\bf Acknowledgements.} I am grateful for stimulating discussions with Erich  Novak and Henryk 
Wo\'{z}niakowski on quantum computing. Parts of this work were done while I was visiting
the Department of Computer Science of the Columbia University, New York, and the Department
of Mathematics of the Hong Kong Baptist University. I thank Joe Traub, Henryk  
Wo\'{z}niakowski and 
Fred Hickernell for their hospitality.

\end{document}